\documentclass[11pt,a4paper]{article}
\usepackage{heppub}
\pdfoutput=1 
\usepackage[svgnames]{xcolor}
\usepackage[normalem]{ulem}

\usepackage{braket,slashed,changepage,placeins,multirow,colortbl,makecell,subcaption,mathtools,array,xspace,enumitem, float,url,todonotes,mathrsfs}
\usepackage[T1]{fontenc}
\usepackage[utf8]{inputenc}

\newcommand{\img}{\mathrm{i}}
\newcommand{\chibar}{\bar{\chi}}

\newcommand{\cB}{\ensuremath{\mathcal{B}}}
\newcommand{\cC}{\ensuremath{\mathcal{C}}}

\newcommand{\cH}{\ensuremath{\mathcal{H}}}
\newcommand{\cL}{\ensuremath{\mathcal{L}}}
\newcommand{\LQCD}{\ensuremath{\Lambda_{\rm QCD}}}
\newcommand{\cM}{\ensuremath{\mathcal{M}}}
\newcommand{\nbar}{\bar{n}}

\newcommand{\cO}{\ensuremath{\mathcal{O}}}

\newcommand{\bP}{\ensuremath{\mathbb{P}}}
\newcommand{\cP}{\ensuremath{\mathcal{P}}}

\newcommand{\qbar}{\bar{q}}

\newcommand{\cT}{\ensuremath{\mathcal{T}}}

\newcommand{\ubar}{\bar{u}}

\newcommand{\cU}{\ensuremath{\mathcal{U}}}

\newcommand{\xbar}{\bar{x}}

\newcommand{\ddslash}{{d\!\!{}^-}}
\newcommand{\IInt}{\int\!\!\!\!\!\int}
\newcommand{\deltaslash}{{\delta\!\!\!{}^-}}
\newcommand{\collapsel}{\Big\{\!\!\!\Big\{}
\newcommand{\collapser}{\Big\}\!\!\!\Big\}}
\newcommand{\scollapsel}{\{\!\!\{}
\newcommand{\scollapser}{\}\!\!\}}

\def\nn{{\nonumber}}

\newcommand{\soft}{\text{\v{S}}}

\DeclareMathOperator*{\sumint}{%
	\mathchoice%
	{\ooalign{$\displaystyle\sum$\cr\hidewidth$\displaystyle\int$\hidewidth\cr}}
	{\ooalign{\raisebox{.14\height}{\scalebox{.7}{$\textstyle\sum$}}\cr\hidewidth$\textstyle\int$\hidewidth\cr}}
	{\ooalign{\raisebox{.2\height}{\scalebox{.6}{$\scriptstyle\sum$}}\cr$\scriptstyle\int$\cr}}
	{\ooalign{\raisebox{.2\height}{\scalebox{.6}{$\scriptstyle\sum$}}\cr$\scriptstyle\int$\cr}}
}

\title{Factorization of diffractive photoproduction}
\preprint{\vbox{%
\hbox{MIT-CTP 6098}
\hbox{LA-UR-26-27745}
}}

\author[a]{Anna Ferdinand,}
\author[b]{Stella T. Schindler,}
\author[a]{and Iain W. Stewart}

\affiliation[a]{Center for Theoretical Physics --  a Leinweber Institute, Massachusetts Institute of Technology\\
Cambridge, MA 02139, USA}
\affiliation[b]{Physics Division P-3, Los Alamos National Laboratory\\
Los Alamos, NM 87545, USA}

\emailAdd{annaferd@mit.edu}
\emailAdd{schindler@lanl.gov}
\emailAdd{iains@mit.edu}

\abstract{
We use effective field theory techniques to factorize the forward scattering dynamics in diffractive photoproduction. Here the photon is on-shell with invariant mass $Q^2=0$, unlike diffractive deep inelastic scattering (dDIS) where $Q^2\gg \Lambda_{\rm QCD}^2$. We show that the forward hadronic matrix element (beam function) in inclusive diffractive photoproduction is the same universal function appearing in dDIS and diffractive proton-proton scattering. This implies that these processes have identical rapidity evolution, which occurs at the amplitude level. This factorization is applicable to ongoing studies at the LHC, such as in ultraperipheral heavy-ion collisions, as well as for planned measurements at the upcoming Electron-Ion Collider.
}

\begin{document}
\maketitle
\allowdisplaybreaks

\section{Introduction}

The vast majority of interactions at the Electron-Ion Collider (EIC) under construction at Brookhaven National Lab will involve \textit{photoproduction}, defined as processes involving the exchange of a nearly on-shell ($Q^2 \approx 0$) photon.
This is due to $Q^{-4}$ growth of the cross section from the photon propagator \cite{H1:2003xoe}.
The same is said to hold true in heavy ion collisions \cite{Busza:2018rrf}.
At HERA, 30-40\% of photoproduction processes were \textit{diffractive}, or forward scattering events with a large rapidity gap in the final state \cite{ZEUS:1994mec, H1:1995hmw, H1:2001txh}.
We anticipate that diffractive photoproduction will also comprise a significant fraction of the total EIC cross section. 
Diffractive processes are critical to longstanding nuclear physics objectives, such as imaging the small-$x$ partonic structure of hadrons
\cite{Frankfurt:2022jns, Hentschinski:2022xnd}, probing nonlinear evolution \cite{Balitsky:1995ub,Jalilian-Marian:1997jhx, Jalilian-Marian:1997qno, Kovchegov:1999yj, Kovner:2000pt, Iancu:2000hn, Iancu:2001ad, Ferreiro:2001qy},  and searching for the onset of gluon saturation \cite{Gribov:1983ivg, Mueller:1985wy, Mueller:1989st, Iancu:2003xm, Gelis:2010nm}.%
\footnote{We separate these references because it has been demonstrated that nonlinear evolution and saturation are not technically connected \cite{Stewart:2023lwz}.}
The DIS Lorentz invariant $x=Q^2/2p\cdot q$ does not appear in photoproduction as $Q^2 \approx 0$, so we preferentially use the terms ``forward'' or ``Regge'' when referring to ``small-$x$'' dynamics later in the paper.
Beyond nuclear physics, diffraction offers a clean search channel for exotic and new particle physics phenomena \cite{Ochs:2013gi, Fichet:2014uka, LHCb:2024smc,Khoze:2001xm, Albrow:2010yb, Knapen:2016moh, Baldenegro:2022kaa, Feng:2022inv, CMS:2023roj, Harland-Lang:2018hmi}, and plays a key role in how high energy cosmic rays scatter through our atmosphere \cite{Knapp:2002vs, 2011ARNPS..61..467E, Grieder2010}. 
Diffractive processes are mediated by the exchange of 
gluons, dressed by quantum fluctutations, whose first principles description remained intractable for over 70 years \cite{Feinberg1956}. This past year, 
an effective field theory (EFT) framework was developed to factorize diffractive deep inelastic scattering (DIS) with $Q^2 \gg \LQCD^2$ \cite{Lee:2025fml} and diffractive proton-proton scattering \cite{Aretz:2026sjb} for the first time. In parallel, there has been significant progress on understanding the structure of amplitudes in the forward (Regge) limit \cite{DelDuca:2011wkl, DelDuca:2011ae, DelDuca:2013ara, Caron-Huot:2013fea, DelDuca:2014cya, Fadin:2016wso, Caron-Huot:2017fxr, Caron-Huot:2017zfo, Fadin:2017nka, Falcioni:2021dgr, Falcioni:2021buo, Falcioni:2020lvv, DelDuca:2021vjq, Caola:2021izf}. Here, we use the EFT approach to factorize diffractive photoproduction, paving the way towards a better understanding of these processes and the unique physical information they encode.

The diffractive photoproduction process $\gamma p \to XY$ is characterized by a forward state $Y$ separated from a spray of central radiation $X$ by a large rapidity gap. We say that this process is \textit{coherent} if the proton stays intact ($Y =p$) and \textit{incoherent} if it dissociates ($Y=$ jet). 
Photoproduction can occur in collisions of electrons ($e$) and protons ($p$), $pp$ collisions, and even ultraperipheral collisions (UPCs) of heavy ions ($A$) \cite{Bertulani:2005ru, Baltz:2007kq, Busza:2018rrf, Klein:2020fmr}. UPCs are relevant for the upcoming High-Luminosity upgrade of the Large Hadron Collider (HL-LHC) at CERN \cite{Azzi:2019yne}, the recently shut-down Relativistic Heavy Ion Collider (RHIC) at Brookhaven \cite{Harrison:2003sb}, and the upcoming Facility for Anti-proton and Ion Research (FAIR) at GSI \cite{CBM:2016kpk}, using beams as heavy as lead and uranium.

HERA took a number of measurements of $ep$ diffractive photoproduction \cite{Newman:2013ada}. To separate diffractive from ungapped events, experimentalists typically  require a rapidity gap of $\Delta\eta \gtrsim 3$. 
For photoproduction, we naturally have $m_e^2 \approx 10^{-7}$ GeV$^2 < Q^2$. A typical measurement integrates over photon virtualities $Q^2 \lesssim 0.01$ GeV$^2$ \cite{H1:2008jdn}.  
For our theory analysis we simply assume that for photoproduction $Q^2\ll \Lambda_{\rm QCD}^2$, in constrast to DIS where $Q^2\gg \Lambda_{\rm QCD}^2$.
HERA probed far larger values of the hadronic momentum transfer $|t|$ in photoproduction than DIS, with measurements up to $\lvert t \rvert \lesssim 30$ GeV$^2$ \cite{H1:2003ksk}.  HERA measurements include both fully inclusive photoproduction \cite{ZEUS:1994mec, H1:1995hmw}, as well as various semi-inclusive and exclusive measurements, such as a coherent forward state \cite{H1:2001txh}; $\rho^0$, $J/\psi$, and $D$ meson tags \cite{H1:2020lzc, H1:2015bxa, H1:2013okq,  ZEUS:2007eio}; and central dijets \cite{H1:2010xdi}. 

Traditional theoretical approaches to diffraction are rooted in parton models and Regge theory. Many Monte Carlo generators rely on the Ingelman-Schlein model \cite{LHCForwardPhysicsWorkingGroup:2016ote}, which describes a diffractive exchange as being mediated by the emission of hypothetical color-singlet Pomeron and Reggeon particles, whose behavior is modeled as a product of diffractive parton distribution functions (dPDFs) and flux factors for Pomerons/Reggeons inside the proton \cite{Ingelman:1984ns}. However, this model does not fully describe HERA data \cite{H1:2007jtx}. Various other models have also been explored over the years, such as soft color interaction (SCI) models \cite{Edin:1995gi, H1:2007jtx}, which also fail to fully capture experimental data. The connection between diffractive photoproduction and saturation has also been explored \cite{Gelis:2001dh, Gelis:2001da, Klein:2002wm, Armesto:2014sma} within the color-glass condensate formalism \cite{McLerran:1993ni, McLerran:1993ka, McLerran:1994vd}. 
Recently, an effective field theory (EFT) of QCD called the soft collinear effective theory (SCET) \cite{Bauer:2000ew,Bauer:2000yr,Bauer:2001ct,Bauer:2001yt,Bauer:2002nz} with Glauber operators \cite{Rothstein:2016bsq} has provided a first-principles factorization of several forward scattering processes, including 2-to-2 amplitudes in the Regge limit \cite{Moult:2022lfy, Gao:2024qsg, Gao:2024fyz}, small-$x$ inclusive DIS \cite{Neill:2023jcd}, diffractive DIS (dDIS) \cite{Lee:2025fml}, as well as elastic, single, and double diffractive $pp$ scattering \cite{Aretz:2026sjb}. 

In this paper, we use SCET to factorize the forward (Regge) physics in diffractive photoproduction. The factorization convolutes a parton distribution function (PDF) for a parton $i$  inside a photon ($f_{i/\gamma}$) with a vacuum matrix element describing the central radiation (soft function $\check S_i$), a proton matrix element describing the forward sector (beam function $B$), and a vacuum matrix element of Wilson lines describing radiation into the gap below the detection threshold (ultrasoft function $U$, relevant for forward jets $Y$); schematically, 
\begin{align}\label{eq:schematic-fact}
   & d\sigma^{\gamma p \to Xp} \propto \sum_i f_{i/\gamma} \otimes_+  \soft_i \otimes_\perp B 
 \,, 
   &&  d\sigma^{\gamma p \to XY} \propto \sum_i f_{i/\gamma} \otimes_+  \soft_i \otimes_\perp B \otimes_{\pm} U 
 \,. 
\end{align}
Note that in photoproduction, there is only one nonzero unpolarized photoproduction structure function $F_\gamma^D$, and the cross section is azimuthally independent, in contrast with the four unpolarized structure functions of diffractive DIS ($F_{L,2,3,4}^D$) \cite{Lee:2025fml}. 
This factorization differs from the Ingelman-Schlein model that is prevalent in Monte Carlo and global fits.

The key object encoding hadronic information in \eq{schematic-fact} is the beam function $B$, which we show is universal across diffractive photoproduction, diffractive DIS, and diffractive $pp$ scattering throughout all of diffractive kinematic phase space. The rapidity evolution of these processes is also universal, and occurs at the amplitude level. 
Importantly, diffractive photoproduction encodes \textit{exactly the same} non-perturbative hadronic matrix element as diffractive DIS, through $B$. 
Because diffractive photoproduction comprises such a dominant portion of experimental data, it is poised to become a uniquely powerful channel for accessing small-$x$ hadron structure and saturation dynamics at the EIC.  

The outline of this paper is as follows. In \sec{formalism}, we introduce the kinematics of diffractive photoproduction and how this naturally leads to an effective field theory description. We use this EFT framework to derive the Regge factorization of diffractive photoproduction forward scattering dynamics in \sec{fact}. We calculate the perturbative component of this factorization formula, the soft function $S_\gamma = \sum_i f_{i/\gamma}\otimes_+ \soft_i$, in \sec{predictions}, and extend our results to photo-diffractive UPCs. Finally, we offer concluding remarks and outlook in \sec{conclusion}. 

\section{Formalism}\label{sec:formalism}

We start in \sec{kinematics} by describing the kinematics of diffractive photoproduction, including the relevant Lorentz invariants and our choice of frame. In \sec{xsec}, we write down the QCD cross section for diffractive photoproduction and define its structure function $F_\gamma^D$. In \sec{pc}, we describe the kinematic constraints that force a process to be diffractive, and show how that leads to natural power expansion parameters. This allows us to determine the appropriate effective field theory for diffractive photoproduction in \sec{eft}. 

\subsection{Kinematics}\label{sec:kinematics}
\begin{figure}[t!]
    \centering
	\includegraphics[width = 2.25 in]{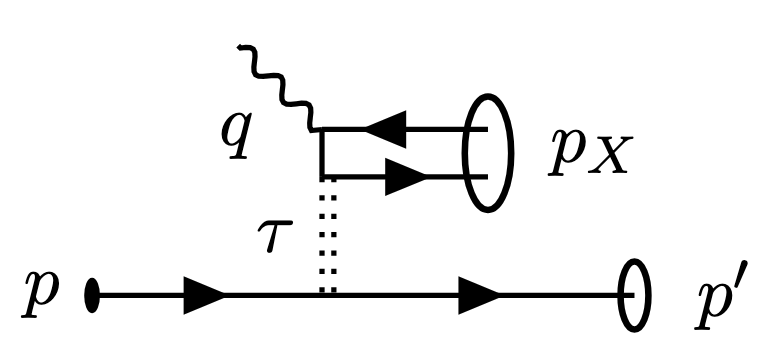} 
	\caption{Diffrative photoproduction can occur in $ep$, $pp$, and heavy ion collisions. The incoming beams emit a nearly on-shell photon with momentum $q$ and a tower of gluons $\tau$. These interact to form a central state $X$ with momentum $p_X$, separated by a large rapidity gap from a forward state $Y$ with momentum $p'$.  $X$ and $Y$ may each consist of a single particle, a collimated jet, or a spray of particles confined to distinct rapidity regions.}\label{fig:diffraction}
\end{figure}

Consider the diffractive scattering of a photon off a proton in \fig{diffraction}. A proton carrying momentum $p$ emits a tower of gluons with total momentum $\tau$. The photon carrying momentum $q$ interacts with the gluon tower $\tau$ to form a hadronic central sector $X$ with overall momentum $p_X = q + \tau$ \cite{Lee:2025fml}. The proton may either stay intact in this collision (coherent) or dissociate into a jet (incoherent), forming a final state $Y$ with overall momentum $p' = p-\tau$. 

From measurements of three incoming and outgoing momenta $p,q,p'$, we can in principle form six linearly independent Lorentz invariants. We  eliminate two of these by taking the photon to be on-shell and using massless proton kinematics 
\begin{align}
	&q^2 = -Q^2 = 0 \,,
	&& p^2 = m_p^2 \approx 0\,,
\end{align}
which leaves us with just four independent invariants. We choose these as three energy scales and one momentum fraction:
\begin{align}\label{eq:invariants}
	&t = \tau^2,
    && W^2 = 2p \cdot q, \:\: 
    && m_Y^2 = (p')^2,
    &&\xi = \frac{\tau \cdot q}{p \cdot q}  \,.
\end{align}
The invariants we use here are different than those of 
diffractive DIS \cite{Lee:2025fml}, which have nonzero $Q^2 \gg \LQCD^2$. For example, in photoproduction we do not have the DIS longitudinal momentum fraction $x= Q^2/(2p\cdot q)$; we will later see that forward scattering is instead controlled by $\xi$.
We also define the dependent invariant
\begin{align}\label{eq:aux-inv}
    &p_X^2 = m_X^2 = t + \xi W^2\,,
\end{align}
for later use.

It is convenient to first consider the $\gamma$-$p$ center of mass frame. We work in 
lightcone coordinates, defined using $p^\pm = (p^t \mp p^z)/2$ and  
\begin{align}
	p^\mu = (p^+,p^-, |p_\perp|) 
	= n\cdot p \frac{\nbar^\mu}{2} + \nbar\cdot p \frac{n^\mu}{2} + p_\perp^{\mu}\,,
\end{align}
where we choose lightlike basis vectors $n^\mu = (0,2,0)$ and $\nbar^\mu = (2,0,0)$, and have $\vec{p}_\perp = (p^x, p^y)$. We can then express dot products as 
\begin{align}
    p\cdot k = (p^+k^- + p^-k^+)/2 + p_\perp \cdot k_\perp\,.
\end{align}
In the center of mass frame, we define $q$ and $p$ to be $\nbar$- and $n$-collinear, respectively:
\begin{align}\label{eq:frame-def}
	&q = W \frac{\bar n}{2} = W(1,0,0)\,,
    &&p = W \frac{n}{2} = W(0,1,0)\,,
\end{align}
where we drop terms of $\cO(m_p^2/W^2)$.  
We can now express the remaining momenta as
\begin{align}\label{eq:cm-momenta}
	\tau &= W\bigg(\frac{t-m_Y^2}{W^2}, \, \xi, \, \frac{1}{W}\sqrt{-t + \xi(t-m_Y^2)} \bigg) \,,
    \\
	p_X &= W\bigg(1 + \frac{t-m_Y^2}{W^2},\, \xi , \, \frac{1}{W}\sqrt{-t + \xi(t-m_Y^2)}  \bigg)\,,
    \nn\\
	p' &= W\bigg(-\frac{t-m_Y^2}{W^2}, \, 1-\xi , \, \frac{1}{W}\sqrt{-t + \xi(t-m_Y^2)} \bigg)\,,
    \nn
\end{align}
up to $\cO(m_p^2/W^2)$ corrections. 

We can immediately deduce a number of constraints on our Lorentz invariants in \eq{invariants}, solely based on positivity conditions. The final state masses $m_X^2$ and $m_Y^2$ must be positive but less than the overall hadronic energy scale $W^2$, which gives the constraint 
\begin{align}
	0 \leq m_Y^2, m_X^2 \leq W^2\,. 
\end{align}
Because $p$, $p^\prime$, and $p_X$ consist of on-shell propagating particles, their longitudinal momentum components must all be positive. This implies that $p^{\prime +} = -\tau^+$ and $p_X^- = \tau^- $ are positive, which means that $t$ is negative:
\begin{align}
	t = \tau^+\tau^- - \vec{\tau}_\perp^2 \leq 0\,. 
\end{align}
From the relation $p^{\prime -} = p^- (1-\xi)$ we also have the constraint
\begin{align}
	0 \leq \xi \leq 1 \,,
\end{align}
from the positivity of $p^{\prime -}$, and $p^-$.

\subsection{Cross section and structure function $F_\gamma^D$}\label{sec:xsec}
The fully differential cross section for diffractive photoproduction is 
\begin{align}\label{eq:starting-point}
	\frac{d^3 \sigma^{\gamma p \to XY}}{d\xi dm_Y^2 dt} = -\frac{\pi^3}{W^2} g_{\mu\nu} W^{\mu\nu} \,,
\end{align}
For the on-shell $Q^2 \approx 0$ dynamics relevant for photoproduction, the photon current projects onto a single tensor structure $\sum_i \epsilon_i^\mu \epsilon_i^{\star\nu} \to -g^{\mu\nu}$. $W^{\mu\nu}$ is the usual hadronic tensor given by
\begin{align}\label{eq:Wmunu}
	W^{\mu\nu}& = \alpha \sumint_X \sumint_Y \delta^4(q+p-p^\prime -p_X) \langle p \lvert J^{\dagger \mu}(0) \rvert YX \rangle  \langle YX \lvert J^\nu(0) \rvert p \rangle \delta(m_Y^2 - p^{\prime 2}) \,,
\end{align}
where $J^\mu = \sum_q e_q \bar \psi_q \gamma^\mu \psi_q$ is a hadronic current, and $q$ indicates quark flavor with corresponding electric charge $e_q$. Note that we absorb a factor of $\alpha$ into the definition of $W^{\mu\nu}$ for later convenience. From \eq{starting-point}, we see that only one structure function appears in photoproduction, 
\begin{align}\label{eq:FiD}
	-g_{\mu \nu} W^{\mu \nu} \equiv F_\gamma^D(\xi, W^2, t, m_Y^2) \,.
\end{align}
This contrasts diffractive DIS, which has an off-shell photon exchange, allowing for four unpolarized structure functions $F_i^D = \cP_{i\mu\nu}W^{\mu\nu}$, with $i = L,2,3,4$ \cite{Lee:2025fml}. In $d=4$, we can express the photoproduction projector $\cP_\gamma^{D\mu\nu} = -g^{\mu\nu}$ in terms of dDIS projectors as 
\begin{align}\label{eq:dis-photo-relation}
    \cP_\gamma^D = \Big[
    -\frac{3}{2x} \cP_L^D + \frac{1}{x}\, \cP_2^D\Big]_{\rm DIS} \,,
\end{align}
where the factors of $1/x$ make the square bracket combination equal to $-g^{\mu\nu}$ without taking the $Q^2\to 0$ limit.
Of course, the dDIS and photoproduction structure functions $F_i^D$ are relevant only in distinct kinematic regimes. Importantly, the photoproduction cross section is azimuthally independent. In contrast, $F_{3,4}^D$ in dDIS give cross section contributions that exhibit a characteristic azimuthal dependence. This distinguishes dDIS from both photoproduction and azimuthally-independent gapped hard DIS \cite{Ee:2026cvm}.
\\[-8 pt]

\noindent\textbf{Photoproduction at the EIC.}
While our analysis centers on forward scattering dynamics in generic photoproduction processes $\gamma p \to XY$, ultimately we are driven by making contact with experimental data. At the EIC, photoproduction occurs when an incoming electron radiates a nearly on-shell photon in a direction collinear to itself. Let the incoming and outgoing electron have momenta $k$ and $k'$. We can now form two more linearly independent Lorentz invariants
\begin{align}
	&s = 2 p \cdot k, 
    && \xbar = \tau \cdot k/p \cdot k, 
\end{align}
where $s$ is the collider energy and $\xbar$ encodes the azimuthal angle of $Y$ relative to the lepton plane in the Breit frame; see \cite{Ee:2026cvm}. 
In DIS kinematics, we describe the $e\to e\gamma$ emission by a leptonic tensor $L^{\mu\nu} = \langle k | J^{\dagger \,\mu} | k' \rangle \langle k' | J^\nu | k\rangle$.
In photoproduction, we typically conceptualize the rate as being proportional to the density of photons inside an electron %
\cite{vonWeizsacker:1934nji, Williams:1934ad}:
\begin{align}\label{eq:photon-pdf}
    d\sigma^{ep \to eXY}= f_{\gamma/e}(y_D)\,
	d\sigma^{\gamma p \to XY}(q,p,p')\,,
\end{align}
Here, $f_{\gamma/e}$  is a parton distribution function (PDF) for photons in the electron and $y_D$ is the standard DIS kinematic variable $y_D = p\cdot q/p \cdot k \approx E_\gamma/E_e$ in the center of mass frame.
We use the shorthand $d\sigma$ to indicate we are fully differential in the relevant Lorentz invariants for both cross sections appearing in this equation.
The literature employs various different models for $f_{\gamma/e}$, such as the modified  Weizsäcker-Williams approximation~\cite{Frixione:1993yw}
\begin{align}
	f_{\gamma/e}(y) \!&=\! \frac{\alpha}{2\pi}\!\!\left[2m_e^2 y \bigg(\frac{1}{Q_{\rm max}^2}- \frac{1}{Q_{\rm min}^2}\bigg) + \frac{1+(1-y)^2}{y}\ln \frac{Q^2_{\rm max}}{Q^2_{\rm min}} \right]
	\!\!\times\!\! \bigg[1 \!+\! \cO\Big(\theta_c^2, \frac{m_e^2}{E^2}\Big)\bigg]. 
\end{align}
Here, we recognize the usual splitting function $[1+(1-y)^2]/y$. This model takes the experimental boundaries for the photoproduction regime to be $Q^2_{\rm min}\approx m_e^2$ and $Q^2_{\rm max}$, and sets $\theta_c\ll 1$ as the maximum angle considered for the scattered electron $k'$.
\\[-8 pt]

\noindent\textbf{Ultraperipheral collisions.} We can also consider diffractive photoproduction in $pp$, $pA$, and $AA$ collisions, 
particularly ultraperipheral collisions (UPCs).
A UPC occurs when two ions pass by each other and interact without direct nuclear overlap. That is, the \textit{impact parameter} $b$, or distance between the centers of the nuclei, is greater than the sum of their radii $b > R_A + R_{A'}$.\footnote{Naturally, we consider the radius along the long (transverse) axis of the ions, which are pancake-shaped in the lab frame when moving at relativistic speeds.} When $b$ is so large as to separate the nuclei significantly, UPCs are predominantly mediated by diffractive photoproduction or direct photonuclear processes \cite{Bertulani:2005ru,Baltz:2007kq}.\footnote{The strong interaction is short range, whereas electromagnetism is long range; thus, in any collision where the nuclei are separated by greater than a few times
the QCD scale ($\sim$fm), at least one ion should emit a photon. Photoproduction generally dominates over double photon exchange at large $b$ as $\alpha_s\gg\alpha_{em}$.} 
We can also consider \emph{grazing UPCs} (or near-UPCs) when $b\gtrsim b_{\rm min}$, for which 
Pomeron exchange can replace photon exchange.

To describe a photoproductive UPC, we simply replace the photon PDF in \eq{photon-pdf} with an appropriate photon flux $n_{\gamma/A}$ from the ion; schematically,
\begin{align}\label{eq:photon-ion-pdf}
    &d\sigma^{AA' \to AXY}_{\rm photo} = n_{\gamma/A}(\omega) \, 	d\sigma^{\gamma A' \to XY}\,.
\end{align}
This flux is approximated in the equivalent photon approximation (EPA) as \cite{Bertulani:2005ru, Baltz:2007kq, Klein:2016yzr, Klein:2020fmr, Frankfurt:2022jns}
\begin{align}\label{eq:epa-flux}
    n_{\gamma/A}(\omega) = \frac{2Z^2\alpha}{\pi\beta^2}\left[\zeta K_0(\zeta)K_1(\zeta) - \frac{\zeta^2}{2}(K_1^2(\zeta) - K_0^2(\zeta) \right]\,.
\end{align}
Here, $K_i$ are modified Bessel functions, $Z$ indicates atomic number,  $\gamma$ and $\beta$ are the boost factor and relativistic velocity, and $\omega$ is the photon energy. 
We also have $\zeta = \omega b_{min} / (\gamma\beta)$, where $b_{min} =R_A + R_{A'}$  for a collision of nuclei with radius $R_{A},R_{A'}$, effectively imposing that we are only considering the UPC case in \eq{epa-flux}.

The literature defines a UPC as a $AA' \to XYZ$ reaction with $b> b_{min}$, which in many key cases is forward, gapped, and thus diffractive \cite{ALICE:2014eof, ALICE:2015mzu, STAR:2023nos, STAR:2025wpi, CMS:2025lsm, CMS:2025oxg, CMS:2026soy,ATLAS:2026cxr, ATLAS:2026sct}. When $b < b_{min}$, we can still have a forward gapped $AA' \to XYZ$ reaction. Conceptually, this separates 
$
    d\sigma^{AA' \to XYZ} = d\sigma_{\rm UPC}(b > b_{min}) + d\sigma_{\rm non\text{-}UPC}(b < b_{min}).
$
Practically, such a separation is not possible, as $b$ is not an observable quantity \cite{Bertulani:2005ru, Baltz:2007kq}. Rather, it is inferred from observables like particle multiplicity \cite{Miller:2007ri} or forward neutron detection in the Zero Degree Calorimeters (ZDCs) \cite{Baltz:2002pp}. 
It is conventional to write
\begin{align}
    d\sigma^{AA' \to XYZ} =  \int d^2b \,N_{\gamma/A}(\omega,b) P(b) \times 	d\sigma^{\gamma A' \to XY}\,,
\end{align}
where $P$ is a probability factor for having no additional hadronic interaction \cite{Klein:2020fmr}. Often the literature takes $P = \theta(b - b_{min})$, which allows us to express the flux in \eq{epa-flux} as
\begin{align}
    n_{\gamma/A} = \int_{b> b_{min}}\!\! d^2b \ N_{\gamma/A} \,.
\end{align}
In the UPC literature, other factors and probability distributions occasionally arise. For example, when using the presence (``$Xn$'') or absence (``$0n$'') of forward neutrons in a ZDC as a proxy for impact parameter in photoproduction measurements, one has to account for the possibility that forward neutrons can also be produced by electromagnetic dissociation (EMD), which is implemented via an additional EMD factor \cite{Cacciari:2025tgr}. 

\subsection{Power counting from forward and gap constraints}\label{sec:pc}

The technique of effective field theory (EFT) entails systematically power expanding the QCD Lagrangian to form an EFT Lagrangian that encodes only the relevant modes for the problem at hand. To factorize the Regge dynamics in diffractive photoproduction, we must determine the appropriate quantity to expand in. From the available Lorentz invariants in \eq{invariants}, we can define four power counting parameters
\begin{align}\label{eq:pc}
	&\lambda^\gamma = \sqrt{\xi},
    &&\lambda^\gamma_t = \frac{\sqrt{-t}}{W\sqrt{\xi}},
    &&\rho = \frac{m_Y}{\sqrt{-t}},
    &&\lambda_\Lambda = \frac{\LQCD}{\sqrt{-t}} \,,
\end{align} 
where we use the notation $\lambda^\gamma, \lambda_t^\gamma$ to differentiate from the power counting parameters $\lambda=Q/\sqrt{s}$ and $\lambda_t=\sqrt{-t}/Q$ of diffractive DIS \cite{Lee:2025fml}.\footnote{Taking the $Q^2\to 0$ limit of $\lambda, \lambda_t$ does not give $\lambda^\gamma, \lambda_t^\gamma$, but we do have the more complicated relationship $(y_D)^{-1/2}\lambda \lambda_t \to  \lambda^\gamma \lambda_t^\gamma$ for $Q^2 \to 0$.}

From the nature of diffractive kinematics \cite{Lee:2025fml, Ee:2026cvm}, we can determine the relative sizes of these power countings. First, we impose forward scattering kinematics, which forces the hadronic momentum transfer $|t|$ to be small relative to the center of mass energy $W^2$:
\begin{align}\label{eq:forward-constraint}
	&-t \ll W^2
    && \implies
    && \lambda^\gamma\lambda^\gamma_t \ll 1\,. 
\end{align}
Next, the positivity constraints on $m_X^2$ and $\xi$ imply that
\begin{align}
	(\lambda^\gamma_t)^2 = 1-\frac{m_X^2}{W^2\xi} \leq 1 \,.
\end{align}
For a process to be diffractive, $X$ and $Y$ must be well separated from one another, requiring that their invariant masses be hierarchically smaller than center of mass energy
\begin{align}\label{eq:collimation-constraint}
	&\LQCD^2 \ll m_X^2 \ll W^2\,,
    &&\implies
    &&\lambda^\gamma \ll 1 \,,
    \nn\\
    &\LQCD^2 \ll m_Y^2 \ll W^2\,,
    &&\implies
    &&(\lambda^\gamma)^2 (\lambda^\gamma_t)^2 \rho^2 \ll 1\,.
\end{align}
$X$ and $Y$ saturate the lower bound when they consist of only a single particle.
In principle, $m_X^2 \ll W^2$ could allow $1-(\lambda^\gamma_t)^2 \ll 1$ without requiring $\lambda^\gamma \ll 1$; however, when combined with \eq{forward-constraint} it forces $\lambda^\gamma \ll 1$.  
Finally, we require a large gap in rapidity $\eta_k = \frac{1}{2}\ln k^-/k^+$ between the central and forward states $X$ and $Y$, which gives the constraint
\begin{align}\label{eq:rapidity-gap}
	&\frac{p^{\prime -}_{\rm Lab}}{p^{\prime +}_{\rm Lab}} \gg \frac{p_{X, \rm Lab}^-}{p_{X, \rm Lab}^+} \,
    &&\implies
    && W^2\frac{1-\xi}{m_Y^2-t} \gg W^2\frac{\xi}{W^2+t-m_Y^2} \,.
\end{align}
This is trivially satisfied given the restrictions in eqs.~\eqref{eq:pc} through \eqref{eq:collimation-constraint}. 
We conclude that diffractive photoproduction requires a power expansion in $\lambda^\gamma \ll 1$, and allows for both $\lambda^\gamma_t \sim 1$ and $\lambda^\gamma_t \ll 1$ scalings, as well as the choice of $\rho \ll 1$, $\rho \sim 1$, and $\rho \gg 1$ scalings.

We can provide a complete description of the forward scattering (Regge) dynamics of the initial state hadron by carrying out the required $\lambda^\gamma \ll 1$ factorization. 
Without loss of generality, we can hold $\lambda_t^\gamma\sim 1,\rho\sim 1$ fixed while carrying out this factorization. 
Considering different hierarchies for these parameters can induce further factorizations, but do not affect the underlying nature of the forward scattering interaction or its factorization. 
It is also beyond the scope of this paper to address the potential power expansion in $\lambda_\Lambda$, which indicates whether the exchange $|t|$ is perturbative or non-perturbative. 
\\[-8 pt]

Working at leading power, we can express the scaling of the momenta in \eq{frame-def} and \eq{cm-momenta} in the $\gamma$-p center of mass frame in terms of $\lambda^\gamma$ and $\lambda^\gamma_t$ as
\begin{align}\label{eq:pc-momenta-original}
     &q = W(1,0,0) \,,
    &&p = W(0,1,0) \,,
    \\
	&\tau \sim W\big( (\lambda^\gamma)^2(\lambda^\gamma_t)^2, (\lambda^\gamma)^2,  \lambda^\gamma\lambda^\gamma_t\big)\,,
	&&p_X \sim W\big(1, (\lambda^\gamma)^2, \lambda^\gamma \lambda^\gamma_t\big)\,,
    \nn\\
	&p^\prime \sim W\big( (\lambda^\gamma)^2(\lambda^\gamma_t)^2, 1, \lambda^\gamma \lambda^\gamma_t\big)\,.
    \nn
\end{align}
For our factorization analysis it is useful to boost to a frame where $\tau$ has Glauber scaling that mediates
$n$-$s$ interactions. (See \sec{eft} below.)
Under a boost $\kappa$, the lightcone coordinates transform as $p^+ \to p^+ /\kappa $ and $p^- \to \kappa p^-$. Boosting by $\kappa = \xi^{-1/2} \sim (\lambda^{\gamma})^{-1}$:
\begin{align}\label{eq:pc-momenta}
    q &= W\sqrt{\xi}\, (1,0,0) \sim W(\lambda^\gamma,0,0)
   \,,  \\
    p &= \frac{W}{\sqrt{\xi}}\, (0,1,0) \sim  W(0,(\lambda^\gamma)^{-1},0) 
   \,, \nn\\ 
	\tau &= W\bigg(\frac{\sqrt{\xi}(t-m_Y^2)}{W^2}, \, \sqrt{\xi}, \, \frac{1}{W}\sqrt{-t + \xi(t-m_Y^2)} \bigg) 
     \sim W\big( (\lambda^\gamma)^3(\lambda^\gamma_t)^2, \lambda^\gamma,  \lambda^\gamma\lambda^\gamma_t\big) 
    \,,
    \nn\\
	p_X &= W\bigg(\sqrt{\xi} + \frac{\sqrt{\xi}(t-m_Y^2)}{W^2},\, \sqrt{\xi} , \, \frac{1}{W}\sqrt{-t + \xi(t-m_Y^2)}  \bigg)
    \sim W\big(\lambda^\gamma, \lambda^\gamma, \lambda^\gamma \lambda^\gamma_t\big)
   \,,
    \nn\\
	p' &= W\bigg(-\frac{\sqrt{\xi}(t-m_Y^2)}{W^2}, \, \frac{1-\xi}{\sqrt{\xi}} , \, \frac{1}{W}\sqrt{-t + \xi(t-m_Y^2)} \bigg)
    \sim W\big( (\lambda^\gamma)^3(\lambda^\gamma_t)^2, (\lambda^\gamma)^{-1}, \lambda^\gamma \lambda^\gamma_t\big)
   \,.
    \nn
\end{align}
We see that $p$ and $p^\prime$ carry $n$-collinear scaling, $p_X$ is soft ($s$), $\tau$ carries Glauber ($G$) scaling, and $q$ is $\nbar$-collinear with respect to $\lambda^\gamma\ll 1$. We thus identify SCET as the relevant EFT for factorizing diffractive photoproduction; we review the salient features of SCET in \sec{eft}. In addition, we note that in kinematic regions with $\lambda^\gamma_t\ll 1$, we have that $p$, $p^\prime$, and $\tau$ are $n$-collinear, $q$ is $\bar n$-collinear, and $p_X$ has hard scaling with respect to $\lambda_t^\gamma$ \cite{Lee:2025fml}.

\begin{figure}
    \centering
        \includegraphics[width =2.25 in]{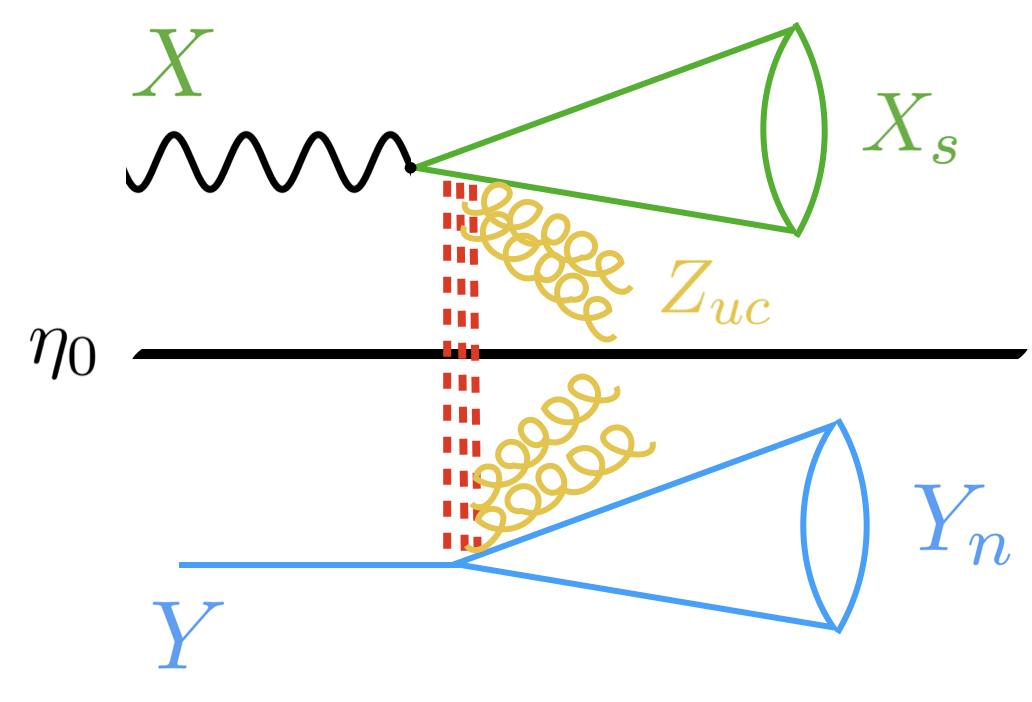}
    \caption{Sectors of diffractive photoproduction, as described in \eq{sectors}. The invariant mass of $X$ includes contributions both from soft radiation $X_s$ (green) as well as ultrasoft-collinear ($uc$) radiation (yellow) above the rapidity cut $\eta_0$ (black line). 
    The invariant mass of $Y$ includes contributions from collinear radiation $Y_n$ (blue) as well as $uc$ radiation below $\eta_0$.
    The red dashed line indicates off-shell Glauber exchanges.}
\end{figure}

In addition to soft and collinear radiation, \refcite{Lee:2025fml} pointed out that it is necessary to build into our formalism the possibility that invisible radiation with energy lower than the experimental detection threshold penetrates the rapidity gap, a phenomenon called quasi-diffraction. This ultrasoft radiation $Z_{uc}$ contributes to measurements of both $X$ and $Y$. To separate it out, we define pure soft and pure collinear states
\begin{align}\label{eq:sectors}
    &X_s = X / Z_{uc} \,,
    &&Y_n = Y / Z_{uc} \,.
\end{align}
We summarize here the pertinent aspects of this radiation; for a complete description of this radiation and its effects we refer the reader to \refcite{Lee:2025fml}. The radiation carries a scaling in the $\gamma$-$p$ center of mass frame of
\begin{align}\label{eq:gap-radiation}
    p_{uc} \sim W(\lambda^{\gamma\,})^2\, \Big((\lambda_t^{\gamma})^2, 1, \lambda^\gamma_t \Big)\,,
\end{align}
This is ultrasoft relative to $(\lambda^\gamma)^2 \ll 1$ but collinear relative to $\lambda^\gamma_t$ in cases when $\lambda_t^\gamma \ll 1$.
It is ultrasoft-collinear ($uc$) in the frame we use for our factorization analysis; see \eq{pc-momenta}.

We impose a hard rapidity cut $\eta_0$ to determine which of the $uc$ particles contributes to which hadronic sector: if a particle with momentum $p_i$ has rapidity $\eta_i<\eta_0$, it contributes to $X$, and we say that it lies in hemisphere $\cH_X$. If its rapidity $\eta_i > \eta_0$, it contributes to $Y$ and lies in hemisphere $\cH_Y$. For convenience, we define 
\begin{align}
    h = e^{-\eta_0}\,.
\end{align}
See \refcite{Lee:2025fml} for expressions relating $\eta_0$ between various frames.

\subsection{SCET operators relevant for photoproduction}\label{sec:eft}

The soft collinear effective theory (SCET) \cite{Bauer:2000ew,Bauer:2000yr,Bauer:2001ct,Bauer:2001yt,Bauer:2002nz} with Glauber operators \cite{Rothstein:2016bsq} is an EFT that provides a rigorous and systematically improvable approach for analyzing diffraction at high energy colliders \cite{Lee:2025fml,Aretz:2026sjb}. The EFT is structured as a power expansion in a parameter $\lambda\ll 1$ (typically the ratio of two energy scales). 
We briefly summarize the salient features of SCET for our analysis here, referring the reader to \refcite{Rothstein:2016bsq, Lee:2025fml} for deeper discussion. 

The scaling of modes in SCET relative to $\lambda$ is as follows 
\begin{center}
\begin{tabular}{ccccc}
    $n$-collinear & $(\lambda^2,1, \lambda)$\,, &\hspace{0.5 in} & 
    $\nbar$-collinear & $(1,\lambda^2, \lambda)\,,$
    \nn\\
    Soft (s) & $(\lambda,\lambda,\lambda)$\,, & & 
    Ultrasoft (us) & $(\lambda^2,\lambda^2,\lambda^2\big)$\,,
    \nn\\
    Hard & $(1,1,0)$\,, & &
    Glauber & $(\lambda^a,\lambda^b, \lambda)$ with $a+b > 2$\,.
\end{tabular}
\end{center}
A hard scattering in SCET arises as a single short-distance hard operator insertion, whereas forward scattering is described by the insertion of longer-distance Glauber exchange contributions. Glaubers can mediate forward scattering between $n$-$\bar n$ and $n$-$s$ sectors. Only the latter is relevant for our analysis here; the pertinent $n$-$s$ Glauber Lagrangian is
\begin{align}\label{eq:two-sector}
	&\cL^{(0)}_{ns,G} = e^{-i x\cdot \cP} \sum_n \cO_{ns}\,, &&  \cO_{ns} = \sum_{ij=q,g}\cO_{ns}^{ij} = \cO_n^{B} \frac{1}{\cP_\perp^2}\cO_s^{B} \,,
\end{align}
in which we define a shorthand notation summing quark and gluon operators
\begin{align}
	\cO^B_n = \cO^{qB}_n + \cO^{gB}_n, && \cO^B_s = \cO_s^{q_n B} + \cO_s^{g_n B}\,.
\end{align}
Here, $\cO_{n}^{iB}$ and $\cO_{s}^{j_n B}$ are collinear and soft operator building blocks, $\cP_\perp^\mu$ is the SCET transverse momentum label operator, and $B,C$ are octet color indices. The $1/\cP_\perp^2$ in the Lagrangian represents the Glauber potential, which mediates interactions between external partons that are instantaneous in longitudinal position and time. The building block operators are fermion and gluon bilinears such as
\begin{align}
    \cO_n^{qB} &= \chibar_n T^B \frac{\slashed{\nbar}}{2}\chi_n\,, 
\end{align}
where $\chi_n$ is a gauge invariant $n$-collinear quark field and $T^B$ is a color matrix. \Refcite{Rothstein:2016bsq} defines the full set of building blocks in detail.

\begin{figure}
    \centering
    \includegraphics[width = 3 in]{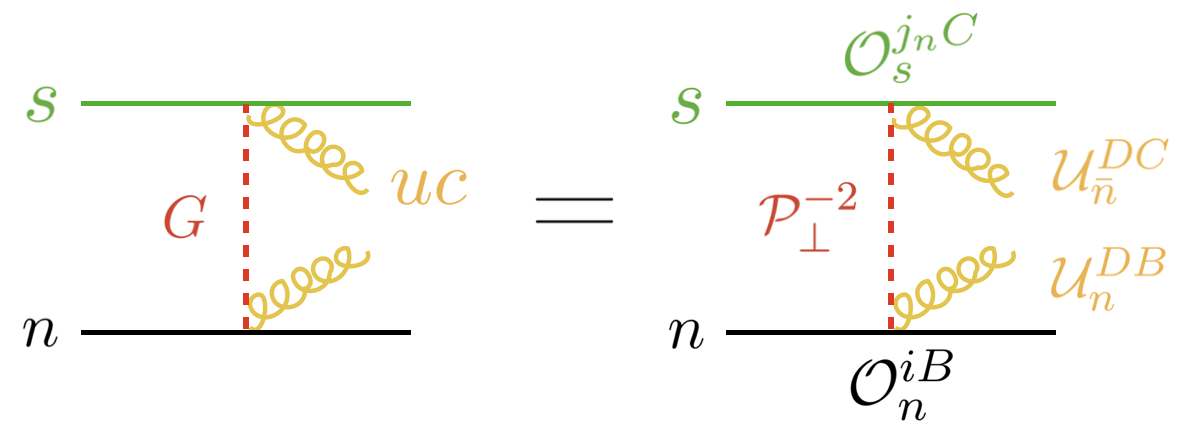}
    \caption{Schematic illustration of Glauber operators for photoproduction in \eq{two-sec-modified}.}\label{fig:ops}
\end{figure}

To describe ultrasoft-collinear radiation that enters a diffractive gap undetected, we use the Bauer-Pirjol-Stewart (BPS) field redefinition \cite{Bauer:2001ct} to decouple ultrasoft-collinear radiation from the $s$ and $n$ quark and gluon fields as
\begin{align}
    &\chi_n \to U_n \chi_n\,,
    &&\cB_{n\perp}^A \to \cU_n^{AB}\cB_{n\perp}^B\,,
\end{align}
where $U_n$ and $\cU_n$ are ultrasoft-collinear Wilson lines in the fundamental and adjoint representations, respectively. As shown in \fig{ops}, this modifies the Glauber operators as
\begin{align}\label{eq:two-sec-modified}
    \cO_{ns} =\cU_n^{DB}\cU_n^{DC}\cO_n^{B} \frac{1}{\cP_\perp^2}\cO_s^{C}\,,
\end{align}
where the Wilson lines obey the relation
\begin{align}
    U_n^\dagger T^A U_n = \cU_n^{AB}T^{B}
\end{align}
and are defined as
\begin{align}
	&  U_n(x) = \bar P \exp\Big( -ig \int_0^\infty\!\! ds\: n\cdot A^A_{uc}(ns + x) T^A \Big) \,,
	\nn\\
	&  \cU_n(x) = \bar P \exp\Big( -ig \int_0^\infty\!\! ds\: n\cdot A^A_{uc}(ns + x) {\cal T}^A \Big) \,.
\end{align}
Here, $T^A$ and $\cT^A$ are fundamental and adjoint color generators, $\bar P$ anti-path-orders the color matrices, and $A_{uc}$ are ultrasoft-collinear gluon fields. See \refcite{Lee:2025fml} for further details. 

\section{Factorization}\label{sec:fact}
We now have all the ingredients we need to factorize inclusive diffractive photoproduction $\gamma p \to XY$. We begin in \sec{factphotoproduction} by carrying out the Regge factorization of diffractive photoproduction, which entails a $\lambda^\gamma \ll 1$ expansion. In \sec{coherent}, we simplify the factorization structure for the case of coherent photoproduction $\gamma p \to Xp$. 
Finally, in \sec{evolution}, we comment on the renormalization group evolution induced by these factorizations.

\subsection{Regge factorization for photoproduction}\label{sec:factphotoproduction}
We start by carrying out the Regge factorization of the forward scattering dynamics in inclusive diffractive photoproduction, which requires us to take $\lambda^\gamma \ll 1$ but is agnostic to the nature of the other power counting parameters. Thus, we take $\lambda^\gamma_t, \rho \sim 1$ here for simplicity; we comment on the effects of further expansions later on. Let us begin with the definition of the photoproduction structure function in \eq{FiD} 
\begin{align}\label{eq:QCDF}
	F_\gamma^D&= -\alpha g_{\mu\nu} \sumint_X\, \sumint_Y \delta^4(q+p-p^\prime -p_X)\, \langle p \lvert J^{\dagger \mu}(0) \rvert YX \rangle  \langle YX \lvert J^\nu(0) \rvert p \rangle\, \delta(m_Y^2 - p^{\prime 2})\,.
\end{align}
Moving to the effective theory, the hadronic current becomes purely soft
\begin{align}\label{eq:factor-J}
    J^\mu 
    \hspace{5pt} \longrightarrow \hspace{5pt}
    J_s^\mu = \sum_q e_q \bar \psi_s^{(q)} \gamma^\mu \psi_s^{(q)}\,,
\end{align}
where we sum over quark flavors $q$. The state $\langle p \lvert$ is purely $n$-collinear, whereas $\lvert XY \rangle$ factorizes into a product state involving soft modes $X_s \in X$, collinear modes $Y_n \in Y$, and ultrasoft-collinear modes $Z_{uc}$ that can belong to $X$ or $Y$. We write these product states as
\begin{align}\label{eq:factor-states}
    &\langle p |  = \langle p_n |  \langle 0_s | \langle 0_{uc}| \,,
    && | XY \rangle = |Z_{uc} \rangle |X_s \rangle | Y_n \rangle  \,, 
\end{align}
To describe the interactions between these modes, we insert the Glauber gluon operator in \eq{two-sec-modified} any number of times. That is, we work in the interaction picture where factorizable pieces of the SCET Lagrangian (pure $n$, $s$, and $uc$ terms) evolve the quantum states, and we explicitly insert the evolution from the non-factorizable Glauber Lagrangian.  

\begin{figure}
    \centering
        \includegraphics[width = 1.75 in]{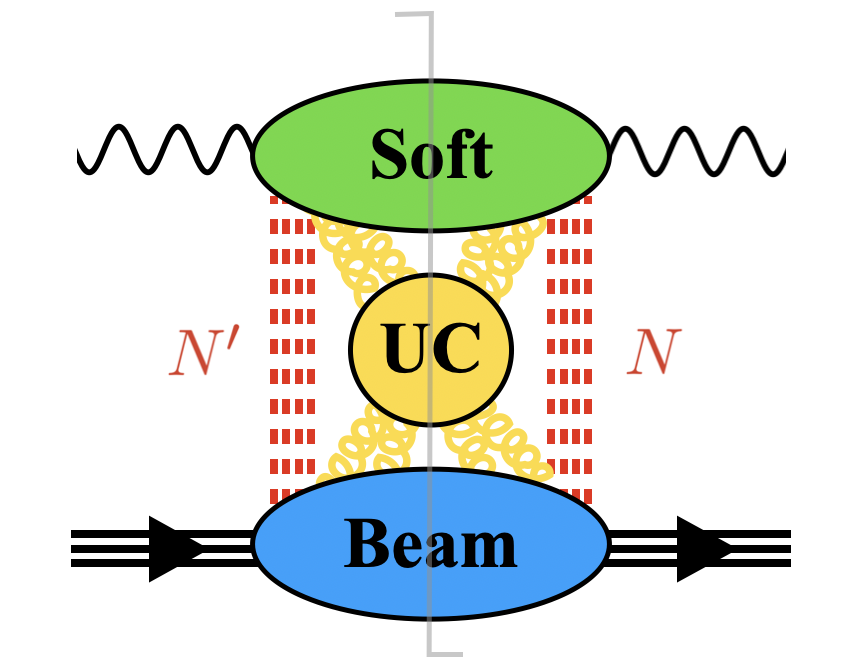}
        \qquad\qquad
        \includegraphics[width = 1.75 in]{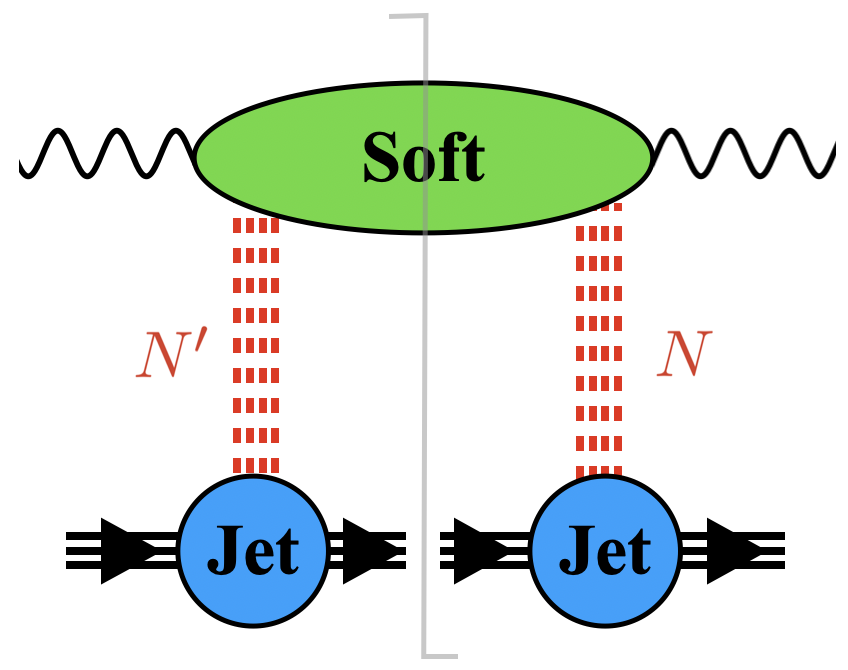}
    \caption{Factorization of incoherent (left, final-state forward jet) and coherent (right, intact proton) diffractive photoproduction, as described in \eqs{fact}{fact-singlet}. Beam/jet (blue) and soft functions (green) are tied together by $N,N'$ Glauber gluons (red) on either side of the cut. Incoherent diffraction also allows radiation into the rapidity gap described by a ultrasoft-collinear ($uc$) function (yellow).}\label{fig:factorization}
\end{figure}

After substituting eqs.~\eqref{eq:factor-J}, \eqref{eq:factor-states}, and any number of insertions of \eqref{eq:two-sec-modified} into $F_\gamma^D$, \eq{QCDF} takes on an operator form that is strikingly similar to that of the diffractive DIS structure functions $F_i^D$ in \refcite{Lee:2025fml}. We can directly parallel the simplification steps for diffractive DIS, to arrive at the fully factorized form of $F_\gamma^D$ depicted in \fig{factorization}: 
\begin{align}\label{eq:fact}
	F_\gamma^D &= 
    \frac{W^2 \zeta^2}{\xi}
    \sum_{N, N^\prime=1}^\infty \sum_{\{R_X\}} \IInt^\perp_{NN'} \int dp_g^+ dp_g^- \,
    B_{(N,N')}^{R_B^NR_B^{N'}} 
    \Big(p^-(p^{\prime +} - p_g^+), \{\tau_{i\perp}, \tau'_{j\perp}\}, t \Big) 
    \nn\\& 
    \times U_{(N,N')}^{R_B^N R_B^{N'}R_S^N R_S^{N'}} 
   \Big(\frac{1}{h\sqrt{\xi}} p_g^+, h \sqrt{\xi} p_g^-, \mu\Big) \:
    S_{\gamma (N,N')}^{R_S^N R_S^{N'}} \Big(q^+ (\tau^- - p_g^-), \{\tau_{i\perp}, \tau_{j\perp}^\prime\}, t\Big)
    \,,
\end{align}
where the beam function $B$ encodes behavior of the forward hadronic sector, the soft function $S_\gamma$ describes the central hadronic sector, and the ultrasoft-collinear function $U$ encodes the undetected gap radiation. We describe these matrix elements in greater detail below. The prefactor $\zeta = (\bar n \cdot p)(n \cdot q)/W^2$ in \eq{fact}  enforces boost invariance of $B$ and $S_\gamma$, and takes the value $\zeta = 1$ in the $\gamma$-$p$ center of mass frame.

\Eq{fact} involves multiple different infinite sums and convolutions, including over the number of Glauber gluons $N,N'$ on each side of the cut; their projections onto color representations $R_X$; and the Glauber momenta $\{\tau_{i\perp},\tau'_{j\perp}\}$. The convolution takes the form
\begin{align} \label{eq:iint_perp}
	  \IInt^\perp_{(N,\,N')} 
	  \!\! \equiv 
	  \frac{(-\img)^{N}(\img)^{N'}}{N!\, N'!}\!
	  \int\! \prod_{i=1}^{N}\! \frac{\ddslash\!^{d'}\!\tau_{i\perp}}{\vec\tau_{i\perp}^{\,2}} 
	   \, \deltaslash^{d'}\!\Bigl(\sum_i \tau_{i\perp} + \tau_\perp\Bigr) 
	  \int\! \prod_{j=1}^{N'} \frac{\ddslash\!^{d'}\!\tau'_{j\perp}}{{\vec\tau_{j\perp}^{\,\prime 2}}} \,  \deltaslash^{d'}\!\Bigl(\sum_j {\tau}^{\prime}_{j\perp}- \tau_\perp\Bigr) 
	 \,,
\end{align}
where $\deltaslash^{d^\prime}(\tau_\perp) = (2\pi)^{d^\prime} \delta^{d^\prime}(\tau_\perp)$ and $\ddslash^{d^\prime} k = d^{d^\prime} k / (2\pi)^{d^\prime}$ for $d^\prime = d - 2$ transverse dimensions.
\\[-8 pt]

The beam function in \eq{fact} consists of a hadronic matrix element that sandwiches $N,N'$ Glauber operators between the forward $n$-collinear states $p$ and $Y_n$:
\begin{align}\label{eq:beam}
    B&_{(N,N')}^{R^N R^{N'}}\big(p^- k^+, \{\tau_{i\perp}, \tau'_{j\perp}\}, t \big)
    = \frac{1}{(\bar{n}\cdot p)^2} \sumint_{Y_n} \int \frac{dv^-}{4\pi p^-}\:  e^{\frac{i}{2}v^- k^+} \\
	&\times 
    \big\langle p \big| \!
   \collapsel\text{\footnotesize $\prod_{i=1}^{N-1}$} 
   \cO_n^{A_i}(v^-\text{\small$\frac{n}{2}$},\tau_{i\perp}) \bar \cO_n^{A_N}(v^-\text{\small$\frac{n}{2}$}) 
   \!\collapser P_N^{R^N}
   \big| Y_{n} \big\rangle 
   \big\langle Y_{n} \big|  P_{N'}^{R^{N'}}
  \collapsel \text{\footnotesize $\prod_{j=1}^{N'-1}$}
  \cO_n^{A'_j}(0,\tau_{j\perp}^\prime) \bar \cO_n^{A'_{N'}}(0) 
  \!\collapser
   \big| p \big\rangle .
   \nn
\end{align}
Here, the notation $\scollapsel...\scollapser$ indicates \textit{Glauber collapse}, a 
feature of SCET that sets all $\pm$ arguments in these Glauber operators to the same value after loop integration \cite{Rothstein:2016bsq}. The $\delta$-function that fixes $\tau = \sum_i \tau_i$ leaves free an integration in each matrix element, which we abbreviate using $\bar O(\tilde y) = \int \ddslash^{d-2}\tau_\perp \cO_n^B(\tilde y, \tau_\perp)$. $P_N^R$ projects an $N$-Glauber exchange onto a color representation $R$.
The presence of the $n$-collinear state $Y_n$ implicitly forces dependence of $B$ on the rapidity cut $\eta_0$.

Importantly, the beam function $B$ is \textit{identical} to the beam functions of diffractive DIS \cite{Lee:2025fml} and elastic, single, and double $pp$ diffractive scattering \cite{Aretz:2026sjb}. We can think of $B$ as playing a roughly analogous role to the PDF in traditional hard scattering, as it is a universal object that appears throughout the factorization of many diffractive processes. We emphasize, however, that $B$ has a far more complicated matrix element structure than a PDF and carries a far more nuanced interpretation. The $B$'s describe any number of perturbative and non-perturbative Glauber gluons which interact with multiple partons inside the proton, whereas a PDF describes a high energy interaction with a single parton. For general kinematics, $B$ has no simple relationship to a PDF, GPD, or any other traditional parton distribution \cite{Lee:2025fml}. Furthermore, unlike inclusive DIS, the soft function $S_\gamma$ may also carry non-perturbative contributions.

One might ask whether $B$ matches onto a simpler distribution in any kinematic regime. This question is beyond the scope of this work. However, we note that the beam function depends the Lorentz invariants $t$ and $m_Y^2$, and thus there are two further expansions that we could consider. First, the parameter $\rho = m_Y/\sqrt{-t}$ controls whether the sector $Y$ is a thin or wide spray of particles. When there are large hierarchies in $\rho$, there will be a further factorization of $B$. Second, we could also consider hierarchies in  $\lambda_\Lambda = \LQCD/\sqrt{-t}$, which controls whether parts of the forward scattering interaction are perturbative or fully nonperturbative.

The ultrasoft collinear function $U$ describes the undetected gap radiation in \eq{gap-radiation}, and exchanges color between $B$ and $S$. $U$ takes the form
\begin{align}\label{eq:ultrasoft}
    U_{(N,N')}^{R_B^N R_B^{N'}R_S^N R_S^{N'}}\!\big(p_H^+,p_H^-\big) &=
    \sumint_{Z_{uc}} \delta\big( p_H^+ -  p_{uc}^{{\cal H}_Y+}\big)\,  
      \delta\big( p_H^- -  p_{uc}^{{\cal H}_X-}\big)
 \\*
 &\hspace{-1cm}
 \times 
  \big\langle 0 \big| P_{N}^{R_S^N} \bar T\,
 \text{\footnotesize $\prod_{i=1}^{N}$}\, \mathbb{U}_{n\nbar}^{A_i B_i}(0) \,
 P_N^{R_B^N} \big| Z_{uc} \big\rangle
 \big\langle Z_{uc} \big| P_{N'}^{R_{S}^{N'}} T\,
 \text{\footnotesize $\prod_{j=1}^{N'}$}\, \mathbb{U}_{n\nbar}^{A'_j B'_j}(0)
  P_{N'}^{R_{B}^{N'}} \big| 0\big\rangle 
  \,,  \nn
\end{align}
where we have used the shorthand notation $\mathbb{U}_{n\bar n}^{BC} = \mathcal{U}_n^{DB} \mathcal{U}_{\bar n}^{DC}$. Note that this ultrasoft function is identical to that of diffractive DIS in Ref.~\cite{Lee:2025fml}. 
Here $p_{uc}^{{\cal H}_Y+}$ and $p_{uc}^{{\cal H}_X-}$ are momentum components in hemisphere regions of the state $Z_{uc}$. Like in dDIS, the function $U$ is only non-trivial when the representations $R$ are not a color singlet, whereas $U_{(N,N')}^{1\, 1\, 1\, 1}(p_H^+, p_H^-) = \delta(p_{H}^+)\delta(p_{H}^-)$.

The soft function is a vacuum matrix element describing formation of the central jet $X_s$ from the photon current and Glauber gluons 
\begin{align}\label{eq:soft}
    S&_{\gamma (N,N')}^{R^N R^{N'}} \big(q^+ k^-, \{\tau_{i\perp}, \tau_{j\perp}^\prime\}, t\big)
    = -\frac{\alpha W^2 \xi}{(n\cdot q)^2}\, g_{\mu\nu}\int[d\tilde y][d\tilde y']\,
    \ddslash^d\!z\: e^{\frac{i}{2}(y^+-y^{\prime+})k^-} e^{iz\cdot q}
    \\
    &\times \sumint_{X_s} 
    \nn \big\langle 0 \big| \bar T\,  J_s^\mu(z) 
  \collapsel\! \text{\footnotesize $\prod_{i=1}^{N}$} 
   \cO_s^{B_i}\!(\tilde y,-\tau_{i\perp}) \!\!\collapser 
  P_N^{R^N} \big| X_s \big\rangle  
   \big\langle X_s \big| P_{N'}^{R^{N'}} T\,
   J_s^\nu(0) 
  \collapsel\! \text{\footnotesize $\prod_{j=1}^{N'}$} 
   \cO_s^{B_j'}\!(\tilde y^{\prime},-\tau_{j\perp}^\prime) \!\!\collapser
   \big| 0 \big\rangle.
\end{align}  
Here $T,\bar T$ indicate (anti-)time ordering operators, which also time-order $B$ through the convolution in \eq{fact}.
(To prove this, one uses that diagrams with crossed Glaubers  vanish \cite{Rothstein:2016bsq}, which enforces a strict shared time-ordering between $B$ and $S$.)
The presence of the state $X_s$ implicitly forces dependence of $S$ on the rapidity cut $\eta_0$. 
This photoproduction $S_\gamma$ and the dDIS soft function $S$ differ by both their functional dependence, required prefactors, and because the $Q^2\to 0$ and infrared limit ($\epsilon\to 0$) do not commute trivially. We calculate $S_\gamma$ perturbatively and discuss this noncommutativity in greater detail in \sec{soft}.

\begin{figure}
	\centering
	\raisebox{1.5 em}{\includegraphics[width = 1.45 in]{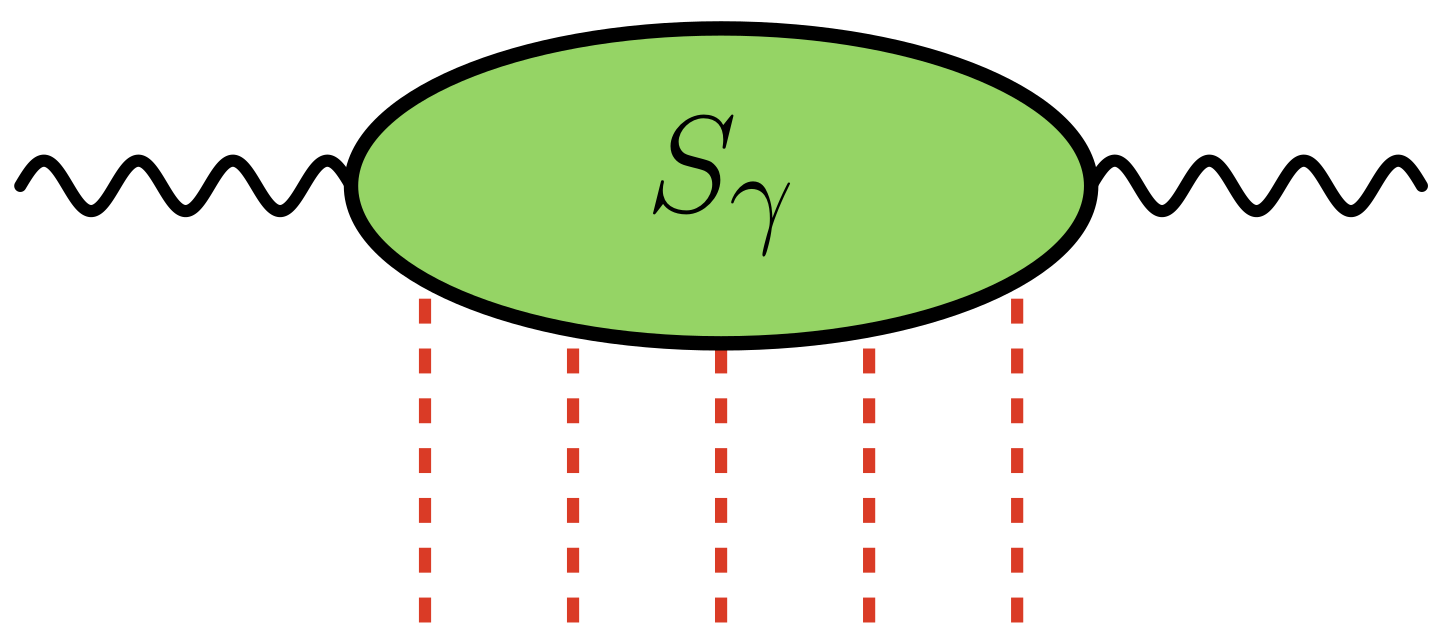}}
	\quad\raisebox{3.5 em}{\LARGE $=$}\quad
	\includegraphics[width = 1.75 in]{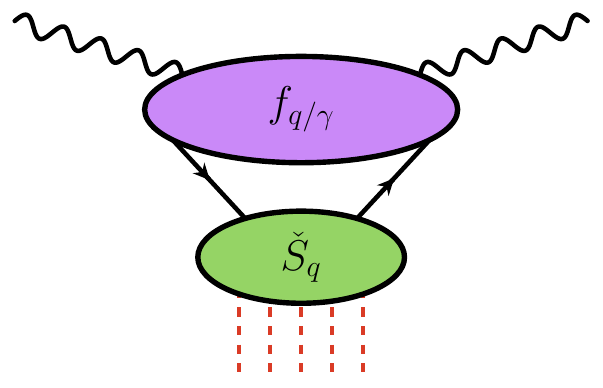}
	\quad\raisebox{3.5 em}{\large{+}}\quad
	\includegraphics[width = 1.75 in]{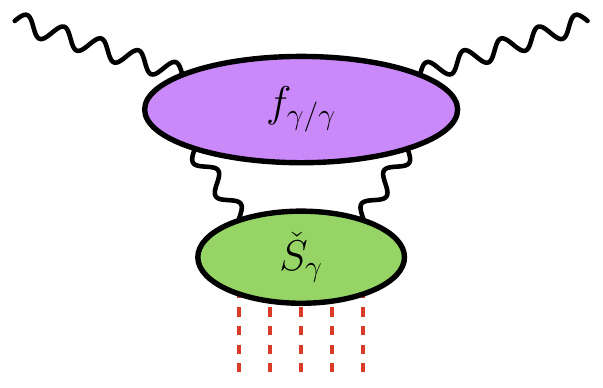}
	\caption{Factorization of the soft function into a photon PDF $f_{i/\gamma}$ and a soft remnant $\check S_i$ in \eq{Sconv}.}
	\label{fig:soft-refact}
\end{figure}

In conventional (hard) photoproduction, we can factorize out a photon PDF $f_{i/\gamma}$ encoding the universal long-distance physics of photon fluctuations from the hadronic sector. 
As the physics of the photon is agnostic to whether the hadronic scattering is hard or diffractive, we can carry out a similar factorization here. For our $\lambda^\gamma \ll 1$ factorization results this acts as a refactorization of our soft function.
Defining $z = -t/(k^- q^+)$ and noting that $0<z < 1$, this factorization in the $\overline {\rm MS}$ scheme is
\begin{align}\label{eq:Sconv}
	S_{\gamma (N,N')}^{R^N R^{N^\prime}}\Big(\frac{-t}{z}, \{\tau_\perp \}, t\Big) 
    &= \Bigg[\sum_i f_{i/\gamma}(y,\mu) \otimes_+  \soft_{i(N,N')}^{R^N R^{N^\prime}}\bigg(\frac{z}{y}, \{\tau_\perp \}, t, \mu\bigg)\Bigg] \times \Big[1 + \cO\Big(\frac{\LQCD}{\sqrt{-t}}\Big) \Big] 
  \nn \\
 & \hspace{-3cm}
= \int_{z}^1 \frac{dy}{y} f_{\gamma/\gamma}(y,\mu)\, 
 \soft_{\gamma (N,N')}^{R^N R^{N^\prime}} \bigg(\frac{z}{y}, \{\tau_\perp \}, t,\mu\bigg) 
 +  2 \int_{z}^1 \frac{dy}{y} f_{q/\gamma}(y,\mu)\, 
 \soft_{q (N,N')}^{R^N R^{N^\prime}} \bigg(\frac{z}{y}, \{\tau_\perp \}, t,\mu \bigg) 
  \nn \\
 & \quad \hspace{-3cm}
 + \ldots
  \,,
\end{align}
where the ellipses involve terms for $f_{g/\gamma}$ that are not needed here.
Here, $f_{i/\gamma}$ is a photon PDF and $\soft_{i}$ is the remnant soft function, as depicted in \fig{soft-refact}.
In the first line, the sum runs over the particles $i \in \{\gamma, q, \bar q, ... \}$ that the photon splits into, and $\otimes_+$ indicates that the convolution runs over $+$ momenta. In the second line, we use that the $q,\qbar$ contributions are equal by $\cC$ symmetry. In \eq{Sconv}, $y$ denotes the $+$ momentum fraction carried by $i$ relative to $\gamma$, and $z = \tau_{\text{eff}}^+/q^+$ is the $+$ momentum fraction carried by Glauber modes, using the derived $+$-momentum $\tau^+_{\text{eff}} = -\tau_\perp^2/\tau^-$ (recalling $\tau^+ = 0$ at leading power in the soft function). Importantly, note that when $p_g^- = 0$ we have $z = -t/W^2 \xi = (\lambda_t^\gamma)^2$. We verify the $z$ dependence of $S_\gamma$ perturbatively in \sec{soft}. 

We index the perturbative expansion of the photon PDF $f_{i/\gamma} = \sum_k f^{(k)}_{i/\gamma}$ by a superscript and provide its value at low orders here
\cite{Klasen:2002xb}
\begin{align}
    &f_{q/\gamma}^{(0)}(y) = 0, 
    && f_{q/\gamma}^{(1)}(y) = -\frac{\alpha}{2\pi} \frac{1}{\epsilon} P_{\gamma\to q}(y) \,, \\
    &f_{\gamma/\gamma}^{(0)}(y) = \delta(1-y),
    && f_{\gamma/\gamma}^{(1)}(y) = 0 \,.
\end{align}
We use $P_{\gamma\to q}$ to denote the $\gamma \to q \bar q$ splitting function
\begin{align}\label{eq:splittingfcn}
	P_{\gamma\to q}(y) = N_c \sum_q e_q^2\, \big[ y^2 + (1-y)^2 \big] \,.
\end{align}
The appearance of the $\epsilon$-pole in $f_{q/\gamma}^{(1)}$ represents the collinear singularity from the splitting. We extract this collinear singularity from our soft function in \sec{soft}. 

A further expansion of the soft function $\check S_i$ can be carried out in the limit $\lambda_t^\gamma = \sqrt{-t}/(W\sqrt{\xi}) \ll 1$.  Because $(\lambda^\gamma_t)^2 = 1-m_X^2/(W^2\xi)$, this limit corresponds to taking large $m_X^2$; for example, considering $X$ to contain dijets~\cite{Klasen:2004qr}. From the scaling of momenta in \eq{pc-momenta}, we anticipate that there may be an additional hard-collinear factorization in this limit, for example with 
$\check S_\gamma = \sum_j \check H_{\gamma}^j \otimes \check S_{c}^j$. Such a structure would parallel that of diffractive DIS for $-t\ll Q^2$ \cite{Lee:2025fml}.  However, because of how the scales enter in our analysis and the presence of the photon PDF, we do not currently have definitive perturbative cross-checks to confirm this factorized structure. Therefore, we leave this for future exploration, including the important questions of whether the photo-diffraction soft-collinear functions $\check S_c^j$ are the same as those of diffractive DIS (which would be necessary to demonstrate the universality of dPDFs for the two processes), and whether the hard coefficient $\check H_{\gamma}^j$ also appears universally in non-diffractive processes.
\\[-8 pt]

Putting these ingredients together the renormalized factorization formula is
\begin{align}\label{eq:factrenorm}
	F_\gamma^D &= 
    \frac{W^2 \zeta^2}{\xi}
    \sum_i \int_z^1 \frac{dy}{y} f_{i/\gamma}(y,\mu)
    \sum_{N, N^\prime=1}^\infty \sum_{\{R_X\}} \IInt^\perp_{NN'} \int dp_g^+ dp_g^- \,
    \\
   &\times  B_{(N,N')}^{R_B^NR_B^{N'}} 
    \Big(m_Y^2 -t - p^- p_g^+, \{\tau_{i\perp}, \tau'_{j\perp}\}, t ,\mu , \frac{\nu \sqrt{\xi}}{W}\Big) 
  \,  U_{(N,N')}^{R_B^N R_B^{N'}R_S^N R_S^{N'}} 
   \Bigl(\frac{1}{h\sqrt{\xi}} p_g^+, h \sqrt{\xi} p_g^-, \mu\Bigr) \:
   \nn\\
   & \times \check S_{i (N,N')}^{R_S^N R_S^{N'}} \bigg( \frac{-t}{y W^2\xi u} ,
 \{\tau_{i\perp}, \tau_{j\perp}^\prime\}, t
    ,\mu , \frac{\nu}{W\sqrt{\xi}} \bigg)
    \,, \nn
\end{align}
where we introduced an invariant mass renormalization scale $\mu$ and rapidity renormalization scale $\nu$.
Here we use the short-hand notation $u \equiv 1 - p_g^-/\tau^-$.

\subsection{Simplification for coherent photoproduction}\label{sec:coherent}

When the proton stays intact with $|Y_n\rangle = |p\rangle$, the factorization structure in \eq{fact} simplifies significantly, as the exchange must be mediated by a color singlet on both sides of the cut. Plugging in color-singlet projectors $R_X=1$ into \eq{ultrasoft}, we find that the ultrasoft Wilson lines collapse, and $U$ reduces to the trivial expression
\begin{align}\label{eq:singlet-u}
    U_{(N,N')}^{1\, 1\, 1\, 1}(p_H^+, p_H^-) = \delta(p_{H}^+)\delta(p_{H}^-)\,,
\end{align}
which at a field-theoretic level shows that the gap is truly empty in the presence of color-singlet exchange~\cite{Lee:2025fml}. 
We no longer have to sum over final forward states $Y_n$, so we can factorize the cross-section level $B$ in \eq{beam} into amplitude-level jet functions $J$ 
\begin{align}
    \lim_{\delta \to 0} 
    \int_{m_p^2-\delta}^{m_p^2+\delta} dm_Y^2 \,
    B_{(N,N')}^{R^NR^{N'}}\!\!\big(p^-p^{\prime +}, \{\tau_{i\perp}, \tau'_{j\perp}\}, t \big)  
    = J_N^{R^N}\!\!(\{\tau_{i\perp}\},t)\,\,
    [ J_{N'}^{R^{N'}}\!\!(\{\tau'_{j\perp}\},t) ]^* \,,
\end{align}
where 
\begin{align}\label{eq:jet}
    J_N^{R^N}(\{\tau'_{j\perp}\},t) &= \frac{1}{\nbar\cdot p}\big\langle p \big| \!
   \collapsel\text{\footnotesize $\prod_{i=1}^{N-1}$} 
   \cO_n^{A_i}(0,\tau_{i\perp}) \bar \cO_n^{A_N}(0) 
   \!\collapser P_N^{R^N}
   \big| p' \big\rangle\,.
\end{align}
Plugging \eqs{singlet-u}{jet} into \eq{fact}, we find that the factorization of coherent diffractive photoproduction takes the form
\begin{align}\label{eq:fact-singlet}
    F_\gamma^{\rm D,coh} &= 
    \frac{W^2\zeta^2}{\xi}
    \sum_i \int_z^1 \frac{dy}{y} f_{i/\gamma}(y,\mu)
    \sum_{N,N
    '=1}^\infty \sum_{R_X=1} \IInt_{NN'}^\perp 
  J_N^{R^N}\!\!\bigg(\{\tau_{i\perp}\},t,\mu, \frac{\nu \sqrt{\xi}}{W} \bigg)
  \nn\\
  & \times
      \big[J_{N'}^{R^{N'}}\!\!\bigg(\{\tau'_{j\perp}\},t, \mu, \frac{\nu \sqrt{\xi}}{W} \bigg)\big]^\star
  \:
    \check S_{i NN'}^{R^N R^{N'}} \!\! 
    \bigg(\frac{-t}{y\xi W^2},\{\tau_{i\perp}, \tau_{j\perp}^\prime\}, t,\mu,  \frac{\nu}{W\sqrt{\xi}} \bigg)
    \,.
\end{align}
Note that the general structure of $S_\gamma$ is unaffected by this simplification, and the general structure of its refactorization into a photon PDF in \eq{Sconv} is preserved as well.

\subsection{Renormalization and evolution}\label{sec:evolution}

\begin{figure}
    \centering
    \includegraphics[height = 1.3 in]{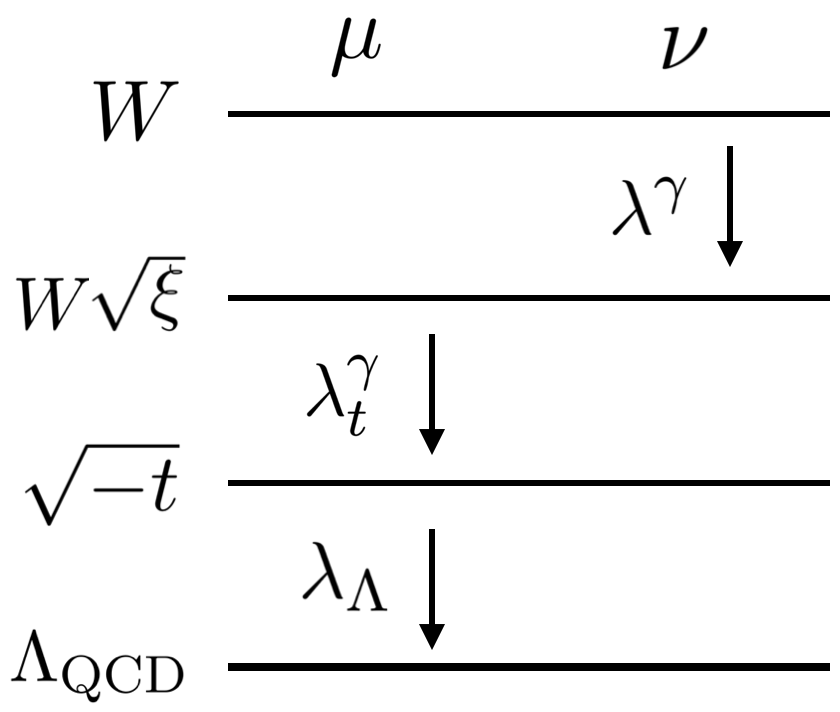}
    \caption{
    Illustration of scale hierarchies in diffractive photoproduction in a kinematic regime with $\lambda^\gamma, \lambda^\gamma_t, \lambda_\Lambda \ll 1$. These can induce a running in virtuality $\mu$ and rapidity $\nu$; see \sec{evolution}. 
     }\label{fig:scales}
\end{figure}

Diffractive photoproduction can involve various hierarchies of scales, as shown in \fig{scales}, each of which can in principle induce renormalization group (RG) evolution. Here, we briefly discuss the evolution we expect to see in diffractive photoproduction, as illustrated in \fig{scales}, relegating detailed calculations of anomalous dimensions and numerical analysis to future work.

In \sec{factphotoproduction}, we carried out the Regge factorization of diffractive photoproduction in $\lambda^\gamma \ll 1$. This induces a rapidity evolution in $\nu$ that is exactly the same as the evolution of diffractive DIS and $pp$ scattering due to the universality of $B$; see \refcite{Aretz:2026sjb} for details. This evolution occurs at the amplitude level, and involves anomalous dimensions $\gamma_{NN'}^{R}$ characterized by three infinite indices: the number of Glaubers $N,N'$ on the left and right sides of the cut, as well as the color representation $R$. Only a small subset of these anomalous dimensions have been calculated thus far \cite{Gao:2024fyz,Gao:2024qsg}, and it would be interesting to explore these in greater detail. Consistency of the EFT factorization theorem implies that $S_\gamma$ has an equal and opposite rapidity evolution to $B$, and hence is the same of $S$ in dDIS, even though these two soft functions are different.

In \sec{factphotoproduction}, we also re-factorized the soft function $S_\gamma$ of diffractive photoproduction into remnant soft functions $\check S_i$ and photon PDFs $f_{i/\gamma}$, which undergo evolution in the photon momentum fraction $z$. The nature of this photon PDF induced evolution is insensitive to Glauber dynamics and is thus identical to that of hard photoproduction; see e.g. \refcite{Klasen:2002xb}.

Virtuality evolution only occurs in the presence of color-nonsinglet quasi-diffractive exchange, the only instance in which we have a nontrivial $U$. The definition of $U$ in diffractive photoproduction is identical to $U$ in diffractive DIS. Thus, diffractive DIS and photoproduction have the same virtuality evolution, which acts as a Sudakov double-logarithmic suppression factor, as discussed in \S5.2 of \refcite{Lee:2025fml}.

\section{Analysis}\label{sec:predictions}

Predictions for diffractive photoproduction require precision knowledge of each of the components appearing in the factorization in \eqs{fact}{fact-singlet}.
The beam ($B$) and jet ($J$) functions both have nonperturbative hadronic contributions, but lattice QCD cannot yet access matrix elements of Glauber operators nor small-$x$ kinematics.
Rather, one may aim to extract these objects from experiment. The ultrasoft-collinear function $U$ is known from diffractive DIS, and induces Sudakov suppression of color-nonsinglet channels \cite{Lee:2025fml}. We calculate the soft function $S_\gamma$ perturbatively in \sec{soft}, in the color octet channel at lowest order in $\alpha_s$. In \sec{scaling}, we make nonperturbative scaling predictions for the structure function $F_\gamma^D$. Finally in \sec{upc}, we discuss implications of our results for ultraperipheral collisions (UPCs) of heavy ions.

\subsection{Perturbative soft function at lowest order in $\alpha_s$}\label{sec:soft}

\begin{figure}
    \centering
        \includegraphics[width = 2 in]{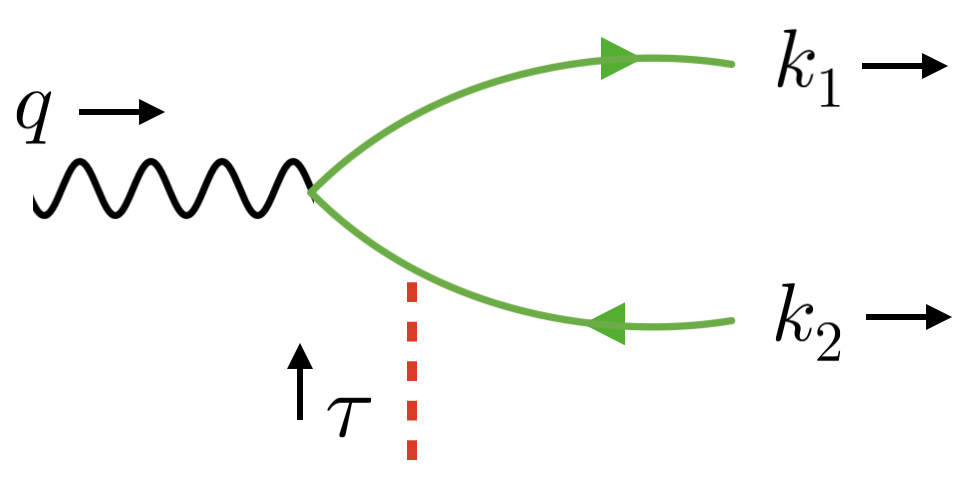}
        \qquad
        \includegraphics[width = 2 in]{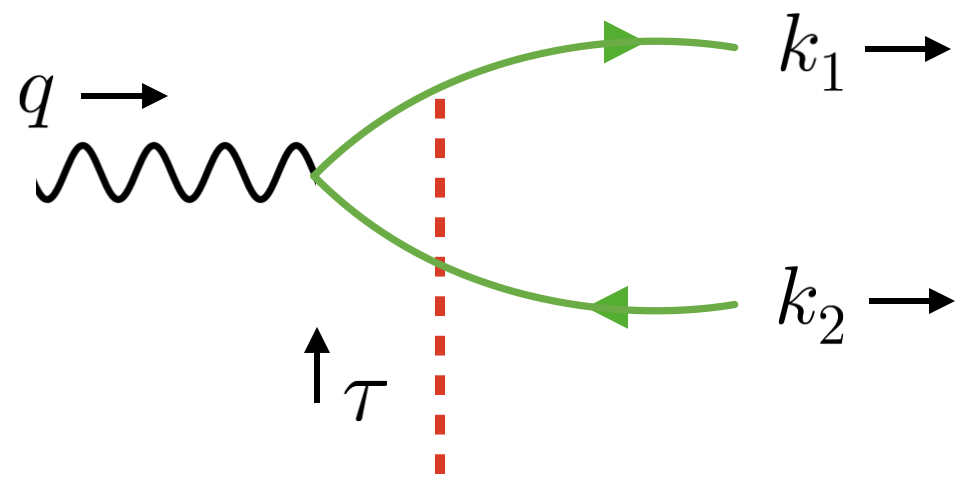}
    \caption{Feynman diagrams for the soft function $S_\gamma$ in \eq{soft} at lowest order in the soft sector ($q\qbar$, green line) and Glauber sector (red dotted line).}\label{fig:soft}
\end{figure}

In this section, we calculate the soft function in the perturbative regime $|t| \gg \LQCD^2$. We work at lowest order in $\alpha_s$, which involves a soft $q\qbar$ emission and a single perturbative Glauber exchange as shown in \fig{soft}.%
\footnote{In principle, one could have a very large number of non-perturbative Glaubers with momenta $\tau_i$ add up to a perturbative overall scale $\tau = \sum \tau_i$. This contribution is suppressed relative to having one perturbative Glauber, due to the low measure of kinematic phase space in which this happens; all such non-perturbative four-vectors $\tau_i$ must line up in the same direction to form the requisite perturbative sum.}
Note that this is the first quasi-diffractive (color-nonsinglet) contribution to photoproduction; the first diffractive (color-singlet) contribution enters with two perturbative Glaubers. 
The relevant amplitude for the diagrams of \fig{soft} directly carries over from diffractive DIS and is provided in eq.~(6.8) of \refcite{Lee:2025fml}. It is
\begin{align}\label{eq:soft-amp-lowest}
    \cM^\mu_{1(8)} = \sum_q e_q T_F \delta^{C B_1}\ubar(k_1) \left[ 
    \gamma^\mu \frac{(\slashed{\tau} - \slashed{k}_2) \frac{\slashed n}{2}}{(\tau - k_2)^2} 
    + \frac{\frac{\slashed n}{2}(\slashed{k}_1 - \slashed{\tau})}{(k_1 - \tau)^2}\gamma^\mu 
    \right]v(-k_2) \,,
\end{align}
where $k_1$ and $k_2$ are the outgoing momenta of the final state quark and antiquark respectively, $\tau$ is the Glauber momentum, and $e_q$ is the quark electric charge. The subscript denotes one Glauber exchange, which is always in the octet channel. We use the following notation to track perturbative order $k$ in the soft sector:
\begin{align}
	S_{\gamma(N,N^\prime)}^{[R_S^N, R_S^{N^\prime}]} =  \sum_{k=0}^{\infty} \bigg(\frac{\alpha_s(\mu)}{2\pi} \bigg)^k  S_{\gamma(N,N')}^{[R_S^N, R_S^{N^\prime}][k]} \,,
\end{align}
Here, we work at leading order $k=0$, corresponding to a tree level soft $q \bar q$ emission. In addition to the factors of $\alpha_s^k$ from the soft sector, there are further factors of $\alpha_s$ from insertions of the Glauber operators. To make this explicit we use the notation
\begin{align} \label{eq:Shat_def}
	S_{\gamma(N,N^\prime)}^{[R_S^N, R_S^{N^\prime}][k]} = \hat S_{\gamma(N,N^\prime)}^{[R_S^N, R_S^{N^\prime}][k]} \frac{1}{(2\pi)^d} \prod_{j=1}^{N+N^\prime} \Big[ 8 \pi \alpha_s(\mu) \mu^{2\epsilon} \iota^\epsilon G(\tau_{j\perp},\mu) \Big] \,,
\end{align}
where $\iota = \exp(\gamma_E)/4\pi$ is the $\overline{\text{MS}}$ factor and $G(\tau_{j\perp},\mu)$ encodes radiative corrections to the Glauber potential.

To calculate $\hat S_{\gamma(1,1)}^{[8,8][0]}$, we include an $\overline{\text{MS}}$ factor also for the electromagnetic coupling $\alpha$, taking $\alpha\to \mu^{2\epsilon} \iota^\epsilon \alpha$. 
We write 
\begin{align}
	\hat S_{\gamma(1,1)}^{[8,8][0]} = \frac{-\alpha \mu^{2\epsilon} \iota^\epsilon W^2 \xi}{(n \cdot q)^2} \int d\Phi_2^d \; g_{\mu\nu} \, \frac{1}{T_F} \cM_{1(8)}^\mu \cM^{\nu\dagger}_{1(8)} \,.
\end{align}
The factor of $1/T_F$ originates from the octet projection, and the two-body phase space factor is
\begin{align}
	\int d \Phi_2^d = \int [d^d k_1]_+ [d^d k_2]_+ (2\pi)^d \delta^d(q+\tau-k_1-k_2) \,,
\end{align}
where $[d^d k_i]_+ = d^d k_i \delta(k_i^2) \theta(k_i^0)/(2\pi)^{d-1}$. 
To evaluate this integral, we work in a frame where the gluon and photon are back-to-back. The momenta and $n^\mu$ are parametrized as
\begin{align}
	&\nn q = (q_0, 0, 0, -q_0), &&\tau = (\tau_0, 0, 0, q_0),\\
	&\nn k_1 = (k_0, k_0\sin\theta, 0, k_0 \cos\theta), &&k_2 = (k_0, -k_0 \sin\theta,0,-k_0 \cos\theta),\\
	&n = (n_0, \lvert n_\perp \rvert \cos \phi, \lvert n_\perp \rvert \sin\phi, n_z) \,,
\end{align}
where the momentum components are
\begin{align}
	&q_0 = \frac{W \sqrt{\xi}}{2}\frac{1}{\sqrt{1+t/W^2\xi}}, && \tau_0 =\frac{W \sqrt{\xi}}{2}\frac{1+2t/W^2 \xi}{\sqrt{1+t/W^2\xi}}\,, &&k_0 = \frac{W\sqrt{\xi}}{2} \sqrt{1+\frac{t}{W^2 \xi}}, 
\end{align}
and the components of $n^\mu$ are
\begin{align}
	&n_0 = \frac{n \cdot q}{W \sqrt{\xi}} \frac{1}{\sqrt{1+t/W^2\xi}}\,, &&n_z = n_0 \bigg(1+\frac{2t}{W^2 \xi}\bigg)\,, &&\lvert n_\perp \rvert = 2 \frac{\sqrt{-t}}{W\sqrt{\xi}} \sqrt{1+\frac{t}{W^2 \xi}} n_0 \,.
\end{align}
In this frame, the phase space factor simplifies to 
\begin{align}
	\int d \Phi_2^d = \frac{(4\pi)^\epsilon}{32\pi^2} \frac{k_0^{-2\epsilon}}{\Gamma(1-\epsilon)} \int \frac{d\cos\theta d\phi}{(1-\cos^2\theta)^\epsilon} \,.
\end{align}
We carry out the angular integrals to obtain the soft function
\begin{align}\label{eq:finalsoftfunction}
	\hat S_{\gamma(1,1)}^{[8,8][0]}= \delta^{AB} T_F \frac{\alpha}{\pi} 
  \frac{-t}{W^2\xi}  \sum_q e_q^2 \Bigg[\hat S_{\gamma, t} \bigg(\frac{1}{\epsilon} + \log\bigg(\frac{\mu^2}{W^2\xi+t}\bigg)\bigg)+\hat S_{\gamma, r} \Bigg] \,,
\end{align}
where the kinematic coefficients are 
\begin{align}\label{eq:softfcncoeffs}
	\hat S_{\gamma, t} = \frac{-2t}{W^2\xi}\Big(1+ \frac{t}{W^2\xi}\Big)-1, \qquad \hat S_{\gamma, r} = \frac{-6t}{W^2\xi}\Big(1+ \frac{t}{W^2\xi}\Big) \,.
\end{align}
We can now explicitly see the refactorization of the soft function into the photon PDF and $\soft_i$ in  \eq{Sconv}. We recognize the functional form of the epsilon pole $\hat S_{\gamma,t}$ as the splitting function $P_{\gamma \to q}(-t/W^2 \xi)$ given in \eq{splittingfcn}. This pole is exactly absorbed by the photon PDF $f_{q/\gamma}$. More concretely, at $\cO{(\alpha)}$ we can expand \eq{Sconv} to yield 
\begin{align}\label{eq:hatSconv1}
	\hat S_\gamma^{(1)}(z) = f_{\gamma/\gamma}^{(0)}(y) \otimes \hat{\soft}_\gamma^{(1)}\bigg(\frac{z}{y}\bigg) + 2f_{q/\gamma}^{(1)}(y) \otimes \hat{\soft}_q^{(0)}\bigg(\frac{z}{y}\bigg) = \hat{\soft}_\gamma^{(1)}(z) + 2N  f_{q/\gamma}^{(1)}(z) \,,
\end{align}
where we use $\hat{\soft}_q^{(0)}(z/y) = N \delta(1-z/y)$ with $N = \delta^{AB} \frac{T_F}{N_c}z$. 
Here we introduce $\hat{\soft}$ as notation for the remnant soft function with the usual hatted factors in \eq{Shat_def} removed.
It is then straightforward to determine $\hat{\soft}_\gamma^{(1)}$ by subtraction of $f_{q/\gamma}^{(1)}$.
After extracting the photon PDF and absorbing the pole into it, we obtain the finite remnant soft function
\begin{align}\label{eq:softfcn_check}
 	\hat{\soft}_{\gamma(1,1)}^{(1)[8,8][0]}= \delta^{AB} T_F \frac{\alpha}{\pi} 
 	\frac{-t}{W^2\xi} \sum_q e_q^2 \Bigg[\hat S_{\gamma, t} \log\bigg(\frac{\mu^2}{W^2\xi+t}\bigg)+\hat S_{\gamma, r} \Bigg] \,.
\end{align}
Finally we note that, up to their overall normalization, the definitions of the soft functions in diffractive DIS and photoproduction are strikingly similar; cf. \eq{soft} here and eq.~(4.25) of \refcite{Lee:2025fml}. However, an important difference is the number of variables, where the dDIS soft function depends on $Q^2$, whereas the one for photoproduction does not.  Taking the $Q^2\to 0$ limit of the dDIS soft function in eq.~(6.22) of  \refcite{Lee:2025fml} projected onto $(-g^{\mu\nu})$ does not yield the soft function for photoproduction because the $\epsilon\to 0$ and $Q^2\to 0$ limits do not commute.  In particular, taking $\epsilon\to 0$ for dDIS gives dependence on a $\tanh^{-1}\big(\sqrt{1-4(-t)Q^2/(\xi^2 W^4)}\,\big)$, which blows up as $Q^2\to 0$ and corresponds to a $1/\epsilon$ pole in the photoproduction soft function.

\begin{figure}
    \centering
    \includegraphics[width = 1.75 in]{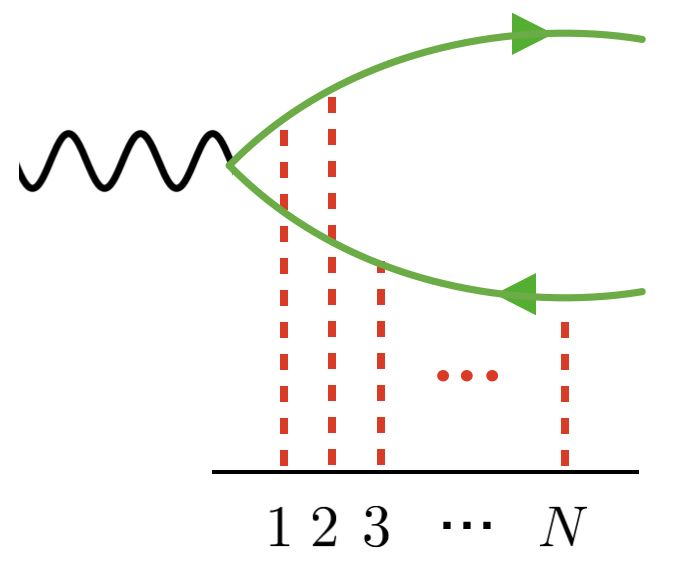}
    \caption{Feynman diagrams for the soft function $S_\gamma$ in \eq{soft} at lowest order in the soft sector ($q\qbar$, green line) and with $N$ Glauber attachments (red dotted lines). Each of the Glauber attachments $i$ may carry a perturbative coupling $\alpha_s(\tau_i) \ll 1$ or a nonperturbative coupling $\alpha_s \sim 1$. This means that an $N$-Glauber diagram can contribute to the cross section at the same order in $\alpha_s$ as an $N=1$ diagram, by having 1 P and ($N-1$) NP Glaubers; see \sec{np-glaubers}.}\label{fig:N-glauber-soft}
\end{figure}

\subsubsection{One perturbative and $N-1$ nonperturbative Glaubers}\label{sec:np-glaubers}

One might na{\"i}vely assume that the diagrams in \fig{soft} are the only contributions to $S_\gamma$ for a $q\qbar$ soft sector that occur at $\cO(\alpha_s)$ in the Glauber sector. However, we also have $N$-Glauber diagrams that occur at $\cO(\alpha_s)$, which have one perturbative Glauber $i=p$ with $\alpha_s(\tau_p^2) \ll 1$ and $N-1$ nonperturbative Glaubers $i \neq p$ with $\alpha_s(\tau_i^2) \sim 1$; see \fig{N-glauber-soft}. Despite the seeming complication from how many diagrams we must consider at each order in $\cO(\alpha_s)$, just as in diffractive DIS \cite{Lee:2025fml}, we still maintain perturbative predictability. 

Let us denote the diffractive photoproduction soft function with 1 perturbative gluon on each side of the cut, along with $N$ and $N'$ nonperturbative gluons on each side respectively, as $S_{\gamma(1,N,1,N')}$. 
It is straightforward to show that
\begin{align}
    S_{\gamma(1,N,1,N')}\Big(\frac{-t}{z},t\Big) = \mathcal{C}_{N,N'} S_{\gamma(1,1)}\Big(\frac{-t}{z},t\Big)  \,,
\end{align}
where $\mathcal{C}_{N,N'}$ involves traces over color matrix commutators, as derived for diffractive DIS in \S6.3.1 of \refcite{Lee:2025fml}; the same result trivially holds true for diffractive photoproduction. Thus, no matter how many Glaubers we add, we do not change the dependence of $S_\gamma$ on its kinematic variables. This suggests a route to perturbative predictability. For example, consider the quasi-diffractive contribution to $F_\gamma^D$ from the color-octet considered here, using $U$ at leading logarithmic order where its color indices are simple.
To separate out the perturbative soft function, we shift the color commutators into the already non-perturbative object $B$ as
\begin{align}
    F_\gamma^{D\,{\rm quasi}} &= U^{LL}  \sum_{N,N'} S_\gamma \otimes_\perp B 
    \nn\\
  & =  U^{LL} \ S_{\gamma(1,1)}  \sum_{N,N'} {\cal C}_{N,N'} \otimes_\perp B
   \times \big[ 1 +{\cal O}(\alpha_s) \big]
\,.
\end{align}
At this order the soft function $S_{\gamma(1,1)}$ is multiplicative, and all perpendicular momentum integrals are internal to the beam function $\tilde B= \sum_{N,N'} {\cal C}_{N,N'}\otimes_\perp B$ that emits a single perturbative Glauber on each side of the cut, thereby enabling predictions for the $z$ dependence in quasi-diffractive photoproduction. 
However, we remark that making this shift likely spoils universality between the beam functions in $ep$ and $pp$ diffraction.

\subsection{Nonperturbative scaling predictions}\label{sec:scaling}

From the factorization formulae of \eqs{fact}{fact-singlet}, we can make nonperturbative predictions for the $\lambda^\gamma$ and $\lambda_t^\gamma$ scalings of the diffractive photoproduction structure function $F_{\gamma}^D$ by using the scaling of the relevant states, prefactors, and SCET operators. Given that the diffractive photoproduction factorization has an analogous structure to that of diffractive DIS and that their beam, soft and ultrasoft-collinear functions have equivalent operator definitions, the diffractive photoproduction scalings in $\lambda^\gamma$ are equivalent to the dDIS scalings in $\lambda$; see Ref.~\cite{Lee:2025fml}.

We can read off the $\lambda_t^\gamma$ scaling from the explicit soft function calculation in \eq{softfcn_check}, from which we find $\soft_\gamma \sim (\lambda_t^\gamma)^2$. This is also in agreement with dDIS; see \eq{dis-photo-relation} and Ref.~\cite{Lee:2025fml}. The $\lambda_t^\gamma$ scalings of the remaining objects ($B$, $U$, integration measures) are equivalent to the dDIS case. Therefore, $F_\gamma^D$ has scaling
\begin{align} \label{eq:FDscale}
	F_\gamma^D \sim
	(\lambda^\gamma)^{-8}(\lambda_t^\gamma)^{-4} \,. 
\end{align} 
This scaling is the overall power counting, but does not account for modifications from anomalous dimensions, which sum $\ln\lambda^\gamma$ and $\ln\lambda_t^\gamma$ factors. Translating to kinematic variables, this scaling is $F_\gamma^D \sim W^4 \xi^{-2}$ for fixed $t$.\footnote{
Note that this classic $\xi^{-2}$ scaling is also  obtained in the Ingelman-Schlein model \cite{Ingelman:1984ns}, in which the proton emits a Pomeron $\bP$, which emits a parton $i$ which then interacts with the photon. Schematically, this gives $F_\gamma^D \sim f_{\bP/p} \times f_{i/\bP} \times \sigma_{\gamma i} $, where $f_{i/j}$ indicates the distribution of a particle $i$ inside $j$. Here, $f_{\bP / p}(\xi, t) \sim \xi^{1-2\alpha_\bP(t)}$ and the cross section is $\sigma_{\gamma i} \sim 1/m_X^2 \sim 1/\xi$ for the $\gamma$-$i$ subprocess. Taking $\alpha_\bP = 1$, this gives an overall $F_\gamma^D\sim \xi^{-2}$ scaling.}
Note that the photoproduction scaling in \eq{FDscale} differs from the dDIS scaling $F_{L,2,3}^D \sim \lambda^{-6} \lambda_t^{-4}$ due to the difference in normalization of their structure functions; see \eq{dis-photo-relation}. 

\subsection{Diffractive photoproduction and UPCs}\label{sec:upc}

Our results on $ep$ and $pp$ diffractive photoproduction naturally generalize to the ultraperipheral collisions (UPCs) of ions discussed in \sec{xsec}. To factorize diffractive photoproduction off an ion in a UPC, we simply replace the proton beam function $B_{p\to Y}$ in \eq{factrenorm} with an analogous ion beam function $B_{A \to Y}$: 
\begin{align}\label{eq:upc-fact}
    d\sigma^{\gamma A \to XY} \propto  \sum_i f_{i/\gamma} \otimes_+ \check S_i \otimes_\perp B_{A\to Y}  \otimes_\pm U\,.
\end{align}
To define $B_{A \to Y}$, we simply replace the proton states $|p\rangle$ in $B_{p \to Y}$ in \eq{beam} with ion states $|A\rangle$.
The factors $f_{i/\gamma}, \check S,$ and $U$ remain identical to $\gamma p \to XY$ photoproduction. Our scaling predictions in \sec{scaling} thus also hold for diffractive photoproduction of ions. Note that this implies that all remarks from earlier parts of this manuscript on what the $\gamma p$ photoproduction factorization reveals about existing models in the literature (dPDFs, Ingelman-Schlein, GPDs/TMDs, etc.) naturally carry over to $\gamma A$ photoproduction as well.

\begin{figure}
    \centering
     \begin{subfigure}{0.3\textwidth}
    	\centering
    	\includegraphics[width = 1.75 in]{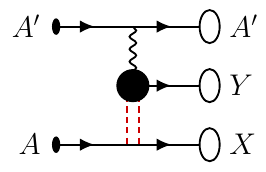}
    	\caption{Photoproduction ($\gamma \bP$)}
    \end{subfigure}
    \begin{subfigure}{0.3\textwidth}
    \centering
        \includegraphics[width = 1.75 in]{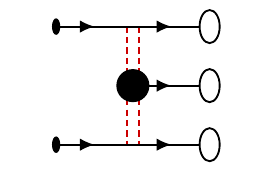}
        \caption{Central diffraction ($\bP \bP$)}
    \end{subfigure}
    \begin{subfigure}{0.3\textwidth}
    \centering
        \includegraphics[width = 1.75 in]{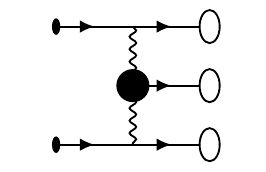}
        \caption{Double photon ($\gamma \gamma$)
        \null}
    \end{subfigure}
    \caption{At small impact parameters, often referred to as near-UPCs or \textit{grazing} collisions, the short-range strong interaction becomes more relevant. This means that double Pomeron exchange or \textit{central diffraction} (middle) can become a sizable background to photoproductive exchanges (left) between ions. Double photon exchange (right) is another possible background.
}\label{fig:double-double}
\end{figure}

To estimate experimental cross sections, we can combine \eq{upc-fact} with the photon flux models in \eqs{photon-pdf}{photon-ion-pdf}. 
As these photon fluxes act multiplicatively, it is interesting to consider ratios of $eA$ and $AA'$ photoproduction under identical kinematics:
\begin{align}\label{eq:ratio-simple}
    \frac{d\sigma^{A'A \to A'XY}_{\rm photo}(k,q,p,p')}{d\sigma^{eA \to eXY}_{\rm photo}(k,q,p,p')}\bigg|_{\text{fixed $t,\xi, W^2 $...}}
    = \frac{n_{\gamma/A'}/y_D}{f_{\gamma/e}} \frac{d\sigma^{\gamma A \to XY}}{d\sigma^{\gamma A \to XY}}
     = \frac{n_{\gamma/A'}[E_\gamma(y_D)]}{y_D\,f_{\gamma/e}(y_D)}\,, 
\end{align}
where the $1/y_D$ factor accounts for a Jacobian change of $d\ln \omega \to dy_D$ between the $e/A'$ photon fluxes. Here $E_\gamma$ denotes the photon energy in the $AA'$ lab frame, as needed to fix $W^2$ in the $\gamma A \to XY$ subprocess.
The EIC and RHIC could allow for tests of various kinematically-matched ratios like this one. 

\paragraph{Diffractive grazing UPCs.} Next, we consider grazing UPCs, in which the nuclei are separated by smaller distances, at which short-range QCD interactions become relevant \cite{Bertulani:2005ru, Baltz:2007kq}. Grazing UPCs are not necessarily photoproductive ($\gamma\bP)$ even at $Q^2 \approx 0$: they could also be mediated by double photon ($\gamma\gamma$) exchange \cite{Klein:2020fmr, Brodsky:1971ud, Budnev:1975poe, ATLAS:2017fur, CMS:2018erd} or double ``Pomeron''/double Glauber/central diffractive ($\bP\bP$) exchange \cite{Cavasinni:1985bi, Roldao:2000ze, Albrow:2010yb, CMS:2023lfr}; see \fig{double-double}. That is, in terms of matrix elements,
\begin{align}\label{eq:light-ion-grazing-contributions}
    {\cal M}^{AA' \to AXY} = {\cal M}^{AA' \to AXY}_{\rm photo} 
     + {\cal M}^{AA' \to AXY}_{\gamma\gamma} + {\cal M}^{AA' \to AXY}_{\bP\bP}  \,. 
\end{align}
The first term is studied in this paper, the second term is purely electromagnetic, while the methods needed for the third term will involve a generalization of \refcite{Aretz:2026sjb} to describe central $pp$ diffraction.
Let us consider the relative magnitudes of these contributions. Photon emission from a heavy ion carries a suppression in the electromagnetic coupling $\alpha_{em}$ and enhancement in the atomic number $Z$, whereas Pomeron/Glauber exchange involves the strong coupling $\alpha_s$ and is enhanced by powers of the atomic mass $A^{k}$.\footnote{The value of $k$ depends on the particular kinematic regime of interest: shadowing, saturation, $\hdots$.} Thus, large impact parameters and heavier ions (such as $^{208}$Pb or $^{238}$U) favor photoproductive processes. Grazing collisions and lighter ions (such as $\alpha$ particles, $^{16}$O, or $^{20}$Ne) could allow for nonnegligible double Pomeron exchange.

Next, we restrict our focus to  light ion grazing collisions in \eq{light-ion-grazing-contributions}.  
As a first test, it would be interesting to check the scaling behavior of the cross section relative to the photoproduction scaling in \sec{scaling}, \eq{FDscale}. Were any significant differences to arise in any kinematic regime, it would motivate making central diffraction predictions. Second, at leading power, diffractive photoproduction can only contribute to the structure function $F_\gamma^D$, so any azimuthal asymmetries between the photon emitter and hadron scattering planes would reflect central diffraction.\footnote{Note that in coherent diffraction with $Y = A$, it is not \textit{a priori} clear which ion emitted the photon. Thus, both cases must be included as terms in the amplitude. In principle, their interference could induce nontrivial azimuthal asymmetries.} Third, taking the ratios of \eq{light-ion-grazing-contributions} with the corresponding $eA$ cross section as in \eq{ratio-simple}, we have
\begin{align}\label{eq:cep-photo-ratio}
    \frac{d\sigma^{A'A \to A' XY}}{d\sigma^{e A \to eXY}_{\rm photo}} &=  \frac{n_{\gamma/A'}[E_\gamma(y_D)]}{y_D f_{\gamma/e}(y_D)} +  \frac{d\sigma^{A'A \to A' XY}_{\bP\bP}}{d\sigma^{e A \to eXY}_{\rm photo}}(k,q,p,p') +  \frac{d\sigma^{A'A \to A' XY}_{\gamma\gamma}}{d\sigma^{e A \to eXY}_{\rm photo}} + \ldots
   \,,
\end{align}
where the ellipses denote cross terms from squaring the amplitudes in \eq{light-ion-grazing-contributions}.
The terms beyond the first (double Pomeron, double photon, and exchanges that differ on either side of the cut) can exhibit nontrivial kinematic dependences on all the momenta $k,q,p,p'$. (We expect that $\gamma\gamma$ exchanges are typically suppressed relative to $\bP\bP$ exchanges.) Thus, any nontrivial kinematic dependences of the cross section on variables besides $y_D$ suggest contributions from central diffraction. 

\section{Conclusions and outlook}\label{sec:conclusion}

In this paper, we derive the Regge factorization of the inclusive diffractive photoproduction process $\gamma p \to X Y$, which enables us to systematically describe its forward scattering dynamics.  We identify that the proton matrix element (beam function) that appears in this factorization is universal to the beam function that appears in diffractive DIS and $pp$ diffraction; however, not all hadronic components (soft functions) are universal. Because the beam function is universal, the rapidity evolution is also universal and occurs at the amplitude level. This universality suggests that diffractive photoproduction is just as powerful a channel as diffractive DIS for accessing target non-perturbative physics encoded in the beam function like small-$x$ hadron structure and gluon saturation dynamics. We calculate the soft function at $q\qbar$ order in soft emissions and for a single Glauber exchange, and discuss its refactorization into a photon PDF and remnant soft function. We provide simple scaling predictions for diffractive photoproduction at the EIC, and extend our results to diffractive photoproduction in ultraperipheral collisions of heavy ions at RHIC and the LHC. 

Our factorization is a first step towards first-principles theoretical control over diffractive photoproduction. 
To enable future precision predictions, it is important to compute the soft function for higher-order soft emissions and more Glauber gluon exchanges, including color-singlet contributions. Such results would also enable exploration of a possible refactorization in the $\lambda_t^\gamma \ll 1$ regime and any further universality between diffractive DIS and photoproduction. Another key direction is extracting the diffractive beam functions from HERA and heavy ion data, and verifying universality across different diffractive processes and collider environments. Furthermore, it would be interesting to extend our results to other types of measurements on diffractive photoproduction processes, such as semi-inclusive and exclusive tags, global observables like energy flow, and studying the impact of nonlinear and saturation dynamics.

\section*{Acknowledgments}
We thank Philipp Aretz, Chris Lee, Ming Liu, and Cesar Da Silva for helpful discussions.
We utilized Claude, Gemini, and/or ChatGPT to assist with literature review, verification of the perturbative soft function calculation, improving our data-free figures, and to do final checks of our manuscript.
S.T.S. thanks Harvard and MIT for hospitality. 
We also thank the Erwin-Schrödinger International Institute for Mathematics and Physics at the University of Vienna for partial support during the Program ``New Paradigms for Harnessing Quantum Field Theory at Colliders'', July 27 – August 28, 2026. 
A.F. and I.S. are supported by the U.S. Department of Energy, Office of Science, Office of Nuclear Physics from DE-SC0011090.
S.T.S. is supported by the LDRD Program of Los Alamos National Laboratory under Project 20240786PRD1. Los Alamos National Laboratory is operated by Triad National Security, LLC, for the National Nuclear Security Administration of the U.S. Department of Energy (Contract Nr. 892332188CNA000001).

\bibliographystyle{JHEP}
\bibliography{bib_AF.bib}

\providecommand{\href}[2]{#2}\begingroup\raggedright\begin{thebibliography}{100}

\bibitem{H1:2003xoe}
{\bf H1} Collaboration, C.~Adloff et~al., {\it {Measurement and QCD analysis of
  neutral and charged current cross-sections at HERA}},  {\em Eur. Phys. J. C}
  {\bf 30} (2003) 1--32, [\href{http://arxiv.org/abs/hep-ex/0304003}{{\tt
  hep-ex/0304003}}].

\bibitem{Busza:2018rrf}
W.~Busza, K.~Rajagopal, and W.~van~der Schee, {\it {Heavy Ion Collisions: The
  Big Picture, and the Big Questions}},  {\em Ann. Rev. Nucl. Part. Sci.} {\bf
  68} (2018) 339--376, [\href{http://arxiv.org/abs/1802.04801}{{\tt
  arXiv:1802.04801}}].

\bibitem{ZEUS:1994mec}
{\bf ZEUS} Collaboration, M.~Derrick et~al., {\it {Measurement of total and
  partial photon proton cross-sections at 180-GeV center-of-mass energy}},
  {\em Z. Phys. C} {\bf 63} (1994) 391--408.

\bibitem{H1:1995hmw}
{\bf H1} Collaboration, S.~Aid et~al., {\it {Measurement of the total
  photon-proton cross-section and its decomposition at 200-GeV center-of-mass
  energy}},  {\em Z. Phys. C} {\bf 69} (1995) 27--38,
  [\href{http://arxiv.org/abs/hep-ex/9509001}{{\tt hep-ex/9509001}}].

\bibitem{H1:2001txh}
{\bf H1} Collaboration, C.~Adloff et~al., {\it {Photoproduction with a leading
  proton at HERA}},  {\em Nucl. Phys. B} {\bf 619} (2001) 3--21,
  [\href{http://arxiv.org/abs/hep-ex/0106070}{{\tt hep-ex/0106070}}].

\bibitem{Frankfurt:2022jns}
L.~Frankfurt, V.~Guzey, A.~Stasto, and M.~Strikman, {\it {Selected topics in
  diffraction with protons and nuclei: past, present, and future}},  {\em Rept.
  Prog. Phys.} {\bf 85} (2022), no.~12 126301,
  [\href{http://arxiv.org/abs/2203.12289}{{\tt arXiv:2203.12289}}].

\bibitem{Hentschinski:2022xnd}
M.~Hentschinski et~al., {\it {White Paper on Forward Physics, BFKL, Saturation
  Physics and Diffraction}},  {\em Acta Phys. Polon. B} {\bf 54} (2023), no.~3
  3--A2, [\href{http://arxiv.org/abs/2203.08129}{{\tt arXiv:2203.08129}}].

\bibitem{Balitsky:1995ub}
I.~Balitsky, {\it {Operator expansion for high-energy scattering}},  {\em Nucl.
  Phys. B} {\bf 463} (1996) 99--160,
  [\href{http://arxiv.org/abs/hep-ph/9509348}{{\tt hep-ph/9509348}}].

\bibitem{Jalilian-Marian:1997jhx}
J.~Jalilian-Marian, A.~Kovner, A.~Leonidov, and H.~Weigert, {\it {The Wilson
  renormalization group for low x physics: Towards the high density regime}},
  {\em Phys. Rev. D} {\bf 59} (1998) 014014,
  [\href{http://arxiv.org/abs/hep-ph/9706377}{{\tt hep-ph/9706377}}].

\bibitem{Jalilian-Marian:1997qno}
J.~Jalilian-Marian, A.~Kovner, A.~Leonidov, and H.~Weigert, {\it {The BFKL
  equation from the Wilson renormalization group}},  {\em Nucl. Phys. B} {\bf
  504} (1997) 415--431, [\href{http://arxiv.org/abs/hep-ph/9701284}{{\tt
  hep-ph/9701284}}].

\bibitem{Kovchegov:1999yj}
Y.~V. Kovchegov, {\it {Small x F(2) structure function of a nucleus including
  multiple pomeron exchanges}},  {\em Phys. Rev. D} {\bf 60} (1999) 034008,
  [\href{http://arxiv.org/abs/hep-ph/9901281}{{\tt hep-ph/9901281}}].

\bibitem{Kovner:2000pt}
A.~Kovner, J.~G. Milhano, and H.~Weigert, {\it {Relating different approaches
  to nonlinear QCD evolution at finite gluon density}},  {\em Phys. Rev. D}
  {\bf 62} (2000) 114005, [\href{http://arxiv.org/abs/hep-ph/0004014}{{\tt
  hep-ph/0004014}}].

\bibitem{Iancu:2000hn}
E.~Iancu, A.~Leonidov, and L.~D. McLerran, {\it {Nonlinear gluon evolution in
  the color glass condensate. 1.}},  {\em Nucl. Phys. A} {\bf 692} (2001)
  583--645, [\href{http://arxiv.org/abs/hep-ph/0011241}{{\tt hep-ph/0011241}}].

\bibitem{Iancu:2001ad}
E.~Iancu, A.~Leonidov, and L.~D. McLerran, {\it {The Renormalization group
  equation for the color glass condensate}},  {\em Phys. Lett. B} {\bf 510}
  (2001) 133--144, [\href{http://arxiv.org/abs/hep-ph/0102009}{{\tt
  hep-ph/0102009}}].

\bibitem{Ferreiro:2001qy}
E.~Ferreiro, E.~Iancu, A.~Leonidov, and L.~McLerran, {\it {Nonlinear gluon
  evolution in the color glass condensate. 2.}},  {\em Nucl. Phys. A} {\bf 703}
  (2002) 489--538, [\href{http://arxiv.org/abs/hep-ph/0109115}{{\tt
  hep-ph/0109115}}].

\bibitem{Gribov:1983ivg}
L.~V. Gribov, E.~M. Levin, and M.~G. Ryskin, {\it {Semihard Processes in QCD}},
   {\em Phys. Rept.} {\bf 100} (1983) 1--150.

\bibitem{Mueller:1985wy}
A.~H. Mueller and J.-w. Qiu, {\it {Gluon Recombination and Shadowing at Small
  Values of x}},  {\em Nucl. Phys. B} {\bf 268} (1986) 427--452.

\bibitem{Mueller:1989st}
A.~H. Mueller, {\it {Small x Behavior and Parton Saturation: A QCD Model}},
  {\em Nucl. Phys. B} {\bf 335} (1990) 115--137.

\bibitem{Iancu:2003xm}
E.~Iancu and R.~Venugopalan, {\em {The Color glass condensate and high-energy
  scattering in QCD}}, pp.~249--3363.
\newblock 3, 2003.
\newblock \href{http://arxiv.org/abs/hep-ph/0303204}{{\tt hep-ph/0303204}}.

\bibitem{Gelis:2010nm}
F.~Gelis, E.~Iancu, J.~Jalilian-Marian, and R.~Venugopalan, {\it {The Color
  Glass Condensate}},  {\em Ann. Rev. Nucl. Part. Sci.} {\bf 60} (2010)
  463--489, [\href{http://arxiv.org/abs/1002.0333}{{\tt arXiv:1002.0333}}].

\bibitem{Stewart:2023lwz}
I.~W. Stewart and V.~Vaidya, {\it {Power counting to saturation}},  {\em Phys.
  Rev. D} {\bf 110} (2024), no.~1 L011504,
  [\href{http://arxiv.org/abs/2305.16393}{{\tt arXiv:2305.16393}}].

\bibitem{Ochs:2013gi}
W.~Ochs, {\it {The Status of Glueballs}},  {\em J. Phys. G} {\bf 40} (2013)
  043001, [\href{http://arxiv.org/abs/1301.5183}{{\tt arXiv:1301.5183}}].

\bibitem{Fichet:2014uka}
S.~Fichet, G.~von Gersdorff, B.~Lenzi, C.~Royon, and M.~Saimpert, {\it
  {Light-by-light scattering with intact protons at the LHC: from Standard
  Model to New Physics}},  {\em JHEP} {\bf 02} (2015) 165,
  [\href{http://arxiv.org/abs/1411.6629}{{\tt arXiv:1411.6629}}].

\bibitem{LHCb:2024smc}
{\bf LHCb} Collaboration, R.~Aaij et~al., {\it {Observation of Exotic
  J/{\ensuremath{\psi}}{\ensuremath{\phi}} Resonant Structure in Diffractive
  Processes in Proton-Proton Collisions}},  {\em Phys. Rev. Lett.} {\bf 134}
  (2025), no.~3 031902, [\href{http://arxiv.org/abs/2407.14301}{{\tt
  arXiv:2407.14301}}].

\bibitem{Khoze:2001xm}
V.~A. Khoze, A.~D. Martin, and M.~G. Ryskin, {\it {Prospects for new physics
  observations in diffractive processes at the LHC and Tevatron}},  {\em Eur.
  Phys. J. C} {\bf 23} (2002) 311--327,
  [\href{http://arxiv.org/abs/hep-ph/0111078}{{\tt hep-ph/0111078}}].

\bibitem{Albrow:2010yb}
M.~G. Albrow, T.~D. Coughlin, and J.~R. Forshaw, {\it {Central Exclusive
  Particle Production at High Energy Hadron Colliders}},  {\em Prog. Part.
  Nucl. Phys.} {\bf 65} (2010) 149--184,
  [\href{http://arxiv.org/abs/1006.1289}{{\tt arXiv:1006.1289}}].

\bibitem{Knapen:2016moh}
S.~Knapen, T.~Lin, H.~K. Lou, and T.~Melia, {\it {Searching for Axionlike
  Particles with Ultraperipheral Heavy-Ion Collisions}},  {\em Phys. Rev.
  Lett.} {\bf 118} (2017), no.~17 171801,
  [\href{http://arxiv.org/abs/1607.06083}{{\tt arXiv:1607.06083}}].

\bibitem{Baldenegro:2022kaa}
C.~Baldenegro, A.~Bellora, S.~Fichet, G.~von Gersdorff, M.~Pitt, and C.~Royon,
  {\it {Searching for anomalous top quark interactions with proton tagging and
  timing detectors at the LHC}},  {\em JHEP} {\bf 08} (2022) 021,
  [\href{http://arxiv.org/abs/2205.01173}{{\tt arXiv:2205.01173}}].

\bibitem{Feng:2022inv}
J.~L. Feng et~al., {\it {The Forward Physics Facility at the High-Luminosity
  LHC}},  {\em J. Phys. G} {\bf 50} (2023), no.~3 030501,
  [\href{http://arxiv.org/abs/2203.05090}{{\tt arXiv:2203.05090}}].

\bibitem{CMS:2023roj}
{\bf CMS, TOTEM} Collaboration, A.~Tumasyan et~al., {\it {A search for new
  physics in central exclusive production using the missing mass technique with
  the CMS detector and the CMS-TOTEM precision proton spectrometer}},  {\em
  Eur. Phys. J. C} {\bf 83} (2023) 827,
  [\href{http://arxiv.org/abs/2303.04596}{{\tt arXiv:2303.04596}}].

\bibitem{Harland-Lang:2018hmi}
L.~A. Harland-Lang, V.~A. Khoze, M.~G. Ryskin, and M.~Tasevsky, {\it {LHC
  Searches for Dark Matter in Compressed Mass Scenarios: Challenges in the
  Forward Proton Mode}},  {\em JHEP} {\bf 04} (2019) 010,
  [\href{http://arxiv.org/abs/1812.04886}{{\tt arXiv:1812.04886}}].

\bibitem{Knapp:2002vs}
J.~Knapp, D.~Heck, S.~J. Sciutto, M.~T. Dova, and M.~Risse, {\it {Extensive air
  shower simulations at the highest energies}},  {\em Astropart. Phys.} {\bf
  19} (2003) 77--99, [\href{http://arxiv.org/abs/astro-ph/0206414}{{\tt
  astro-ph/0206414}}].

\bibitem{2011ARNPS..61..467E}
R.~{Engel}, D.~{Heck}, and T.~{Pierog}, {\it {Extensive Air Showers and
  Hadronic Interactions at High Energy}},  {\em Annual Review of Nuclear and
  Particle Science} {\bf 61} (Nov., 2011) 467--489.

\bibitem{Grieder2010}
P.~K. Grieder, {\it Hadronic interactions and cascades},  in {\em Extensive Air
  Showers: High Energy Phenomena and Astrophysical Aspects -- A Tutorial,
  Reference Manual and Data Book}, pp.~77--146.
\newblock Springer Berlin Heidelberg, Berlin, Heidelberg, 2010.

\bibitem{Feinberg1956}
E.~L. Feinberg and I.~Pomeran{\v{c}}uk, {\it High energy inelastic diffraction
  phenomena},  {\em Il Nuovo Cimento (1955-1965)} {\bf 3} (Apr, 1956) 652--671.

\bibitem{Lee:2025fml}
K.~Lee, S.~T. Schindler, and I.~W. Stewart, {\it {Effective field theory
  factorization for diffraction}},  {\em JHEP} {\bf 11} (2025) 157,
  [\href{http://arxiv.org/abs/2508.10231}{{\tt arXiv:2508.10231}}].

\bibitem{Aretz:2026sjb}
P.~B. Aretz, K.~Lee, S.~T. Schindler, and I.~W. Stewart, {\it {Factorization of
  elastic, single, and double diffractive $pp$ scattering}},
  \href{http://arxiv.org/abs/2607.07788}{{\tt arXiv:2607.07788}}.

\bibitem{DelDuca:2011wkl}
V.~Del~Duca, C.~Duhr, E.~Gardi, L.~Magnea, and C.~D. White, {\it {An infrared
  approach to Reggeization}},  {\em Phys. Rev. D} {\bf 85} (2012) 071104,
  [\href{http://arxiv.org/abs/1108.5947}{{\tt arXiv:1108.5947}}].

\bibitem{DelDuca:2011ae}
V.~Del~Duca, C.~Duhr, E.~Gardi, L.~Magnea, and C.~D. White, {\it {The Infrared
  structure of gauge theory amplitudes in the high-energy limit}},  {\em JHEP}
  {\bf 12} (2011) 021, [\href{http://arxiv.org/abs/1109.3581}{{\tt
  arXiv:1109.3581}}].

\bibitem{DelDuca:2013ara}
V.~Del~Duca, G.~Falcioni, L.~Magnea, and L.~Vernazza, {\it {High-energy QCD
  amplitudes at two loops and beyond}},  {\em Phys. Lett. B} {\bf 732} (2014)
  233--240, [\href{http://arxiv.org/abs/1311.0304}{{\tt arXiv:1311.0304}}].

\bibitem{Caron-Huot:2013fea}
S.~Caron-Huot, {\it {When does the gluon reggeize?}},  {\em JHEP} {\bf 05}
  (2015) 093, [\href{http://arxiv.org/abs/1309.6521}{{\tt arXiv:1309.6521}}].

\bibitem{DelDuca:2014cya}
V.~Del~Duca, G.~Falcioni, L.~Magnea, and L.~Vernazza, {\it {Analyzing
  high-energy factorization beyond next-to-leading logarithmic accuracy}},
  {\em JHEP} {\bf 02} (2015) 029, [\href{http://arxiv.org/abs/1409.8330}{{\tt
  arXiv:1409.8330}}].

\bibitem{Fadin:2016wso}
V.~S. Fadin, {\it {Particularities of the NNLLA BFKL}},  {\em AIP Conf. Proc.}
  {\bf 1819} (2017), no.~1 060003, [\href{http://arxiv.org/abs/1612.04481}{{\tt
  arXiv:1612.04481}}].

\bibitem{Caron-Huot:2017fxr}
S.~Caron-Huot, E.~Gardi, and L.~Vernazza, {\it {Two-parton scattering in the
  high-energy limit}},  {\em JHEP} {\bf 06} (2017) 016,
  [\href{http://arxiv.org/abs/1701.05241}{{\tt arXiv:1701.05241}}].

\bibitem{Caron-Huot:2017zfo}
S.~Caron-Huot, E.~Gardi, J.~Reichel, and L.~Vernazza, {\it {Infrared
  singularities of QCD scattering amplitudes in the Regge limit to all
  orders}},  {\em JHEP} {\bf 03} (2018) 098,
  [\href{http://arxiv.org/abs/1711.04850}{{\tt arXiv:1711.04850}}].

\bibitem{Fadin:2017nka}
V.~S. Fadin and L.~N. Lipatov, {\it {Reggeon cuts in QCD amplitudes with
  negative signature}},  {\em Eur. Phys. J. C} {\bf 78} (2018), no.~6 439,
  [\href{http://arxiv.org/abs/1712.09805}{{\tt arXiv:1712.09805}}].

\bibitem{Falcioni:2021dgr}
G.~Falcioni, E.~Gardi, N.~Maher, C.~Milloy, and L.~Vernazza, {\it
  {Disentangling the Regge Cut and Regge Pole in Perturbative QCD}},  {\em
  Phys. Rev. Lett.} {\bf 128} (2022), no.~13 132001,
  [\href{http://arxiv.org/abs/2112.11098}{{\tt arXiv:2112.11098}}].

\bibitem{Falcioni:2021buo}
G.~Falcioni, E.~Gardi, N.~Maher, C.~Milloy, and L.~Vernazza, {\it {Scattering
  amplitudes in the Regge limit and the soft anomalous dimension through four
  loops}},  {\em JHEP} {\bf 03} (2022) 053,
  [\href{http://arxiv.org/abs/2111.10664}{{\tt arXiv:2111.10664}}].

\bibitem{Falcioni:2020lvv}
G.~Falcioni, E.~Gardi, C.~Milloy, and L.~Vernazza, {\it {Climbing three-Reggeon
  ladders: four-loop amplitudes in the high-energy limit in full colour}},
  {\em Phys. Rev. D} {\bf 103} (2021) L111501,
  [\href{http://arxiv.org/abs/2012.00613}{{\tt arXiv:2012.00613}}].

\bibitem{DelDuca:2021vjq}
V.~Del~Duca, R.~Marzucca, and B.~Verbeek, {\it {The gluon Regge trajectory at
  three loops from planar Yang-Mills theory}},  {\em JHEP} {\bf 01} (2022) 149,
  [\href{http://arxiv.org/abs/2111.14265}{{\tt arXiv:2111.14265}}].

\bibitem{Caola:2021izf}
F.~Caola, A.~Chakraborty, G.~Gambuti, A.~von Manteuffel, and L.~Tancredi, {\it
  {Three-Loop Gluon Scattering in QCD and the Gluon Regge Trajectory}},  {\em
  Phys. Rev. Lett.} {\bf 128} (2022), no.~21 212001,
  [\href{http://arxiv.org/abs/2112.11097}{{\tt arXiv:2112.11097}}].

\bibitem{Bertulani:2005ru}
C.~A. Bertulani, S.~R. Klein, and J.~Nystrand, {\it {Physics of
  ultra-peripheral nuclear collisions}},  {\em Ann. Rev. Nucl. Part. Sci.} {\bf
  55} (2005) 271--310, [\href{http://arxiv.org/abs/nucl-ex/0502005}{{\tt
  nucl-ex/0502005}}].

\bibitem{Baltz:2007kq}
A.~J. Baltz et~al., {\it {The Physics of Ultraperipheral Collisions at the
  LHC}},  {\em Phys. Rept.} {\bf 458} (2008) 1--171,
  [\href{http://arxiv.org/abs/0706.3356}{{\tt arXiv:0706.3356}}].

\bibitem{Klein:2020fmr}
S.~Klein and P.~Steinberg, {\it {Photonuclear and Two-photon Interactions at
  High-Energy Nuclear Colliders}},  {\em Ann. Rev. Nucl. Part. Sci.} {\bf 70}
  (2020) 323--354, [\href{http://arxiv.org/abs/2005.01872}{{\tt
  arXiv:2005.01872}}].

\bibitem{Azzi:2019yne}
P.~Azzi et~al., {\it {Report from Working Group 1}: {Standard Model Physics at
  the HL-LHC and HE-LHC}},  {\em CERN Yellow Rep. Monogr.} {\bf 7} (2019)
  1--220, [\href{http://arxiv.org/abs/1902.04070}{{\tt arXiv:1902.04070}}].

\bibitem{Harrison:2003sb}
M.~Harrison, T.~Ludlam, and S.~Ozaki, {\it {RHIC project overview}},  {\em
  Nucl. Instrum. Meth. A} {\bf 499} (2003) 235--244.

\bibitem{CBM:2016kpk}
{\bf CBM} Collaboration, T.~Ablyazimov et~al., {\it {Challenges in QCD matter
  physics --The scientific programme of the Compressed Baryonic Matter
  experiment at FAIR}},  {\em Eur. Phys. J. A} {\bf 53} (2017), no.~3 60,
  [\href{http://arxiv.org/abs/1607.01487}{{\tt arXiv:1607.01487}}].

\bibitem{Newman:2013ada}
P.~Newman and M.~Wing, {\it {The Hadronic Final State at HERA}},  {\em Rev.
  Mod. Phys.} {\bf 86} (2014), no.~3 1037,
  [\href{http://arxiv.org/abs/1308.3368}{{\tt arXiv:1308.3368}}].

\bibitem{H1:2008jdn}
{\bf H1} Collaboration, F.~D. Aaron et~al., {\it {Measurement of Diffractive
  Scattering of Photons with Large Momentum Transfer at HERA}},  {\em Phys.
  Lett. B} {\bf 672} (2009) 219--226,
  [\href{http://arxiv.org/abs/0810.3096}{{\tt arXiv:0810.3096}}].

\bibitem{H1:2003ksk}
{\bf H1} Collaboration, A.~Aktas et~al., {\it {Diffractive photoproduction of
  $J/\psi$ mesons with large momentum transfer at HERA}},  {\em Phys. Lett. B}
  {\bf 568} (2003) 205--218, [\href{http://arxiv.org/abs/hep-ex/0306013}{{\tt
  hep-ex/0306013}}].

\bibitem{H1:2020lzc}
{\bf H1} Collaboration, V.~Andreev et~al., {\it {Measurement of Exclusive
  $\pi^{+}\pi^{-}$ and $\rho^0$ Meson Photoproduction at HERA}},  {\em Eur.
  Phys. J. C} {\bf 80} (2020), no.~12 1189,
  [\href{http://arxiv.org/abs/2005.14471}{{\tt arXiv:2005.14471}}].

\bibitem{H1:2015bxa}
{\bf H1} Collaboration, V.~Andreev et~al., {\it {Exclusive $\rho^0$ meson
  photoproduction with a leading neutron at HERA}},  {\em Eur. Phys. J. C} {\bf
  76} (2016), no.~1 41, [\href{http://arxiv.org/abs/1508.03176}{{\tt
  arXiv:1508.03176}}].

\bibitem{H1:2013okq}
{\bf H1} Collaboration, C.~Alexa et~al., {\it {Elastic and Proton-Dissociative
  Photoproduction of J/psi Mesons at HERA}},  {\em Eur. Phys. J. C} {\bf 73}
  (2013), no.~6 2466, [\href{http://arxiv.org/abs/1304.5162}{{\tt
  arXiv:1304.5162}}].

\bibitem{ZEUS:2007eio}
{\bf ZEUS} Collaboration, S.~Chekanov et~al., {\it {Diffractive photoproduction
  of D*+-(2010) at HERA}},  {\em Eur. Phys. J. C} {\bf 51} (2007) 301--315,
  [\href{http://arxiv.org/abs/hep-ex/0703046}{{\tt hep-ex/0703046}}].

\bibitem{H1:2010xdi}
{\bf H1} Collaboration, F.~D. Aaron et~al., {\it {Diffractive Dijet
  Photoproduction in ep Collisions at HERA}},  {\em Eur. Phys. J. C} {\bf 70}
  (2010) 15--37, [\href{http://arxiv.org/abs/1006.0946}{{\tt
  arXiv:1006.0946}}].

\bibitem{LHCForwardPhysicsWorkingGroup:2016ote}
{\bf LHC Forward Physics Working Group} Collaboration, K.~Akiba et~al., {\it
  {LHC Forward Physics}},  {\em J. Phys. G} {\bf 43} (2016) 110201,
  [\href{http://arxiv.org/abs/1611.05079}{{\tt arXiv:1611.05079}}].

\bibitem{Ingelman:1984ns}
G.~Ingelman and P.~E. Schlein, {\it {Jet Structure in High Mass Diffractive
  Scattering}},  {\em Phys. Lett. B} {\bf 152} (1985) 256--260.

\bibitem{H1:2007jtx}
{\bf H1} Collaboration, A.~Aktas et~al., {\it {Tests of QCD factorisation in
  the diffractive production of dijets in deep-inelastic scattering and
  photoproduction at HERA}},  {\em Eur. Phys. J. C} {\bf 51} (2007) 549--568,
  [\href{http://arxiv.org/abs/hep-ex/0703022}{{\tt hep-ex/0703022}}].

\bibitem{Edin:1995gi}
A.~Edin, G.~Ingelman, and J.~Rathsman, {\it {Soft color interactions as the
  origin of rapidity gaps in DIS}},  {\em Phys. Lett. B} {\bf 366} (1996)
  371--378, [\href{http://arxiv.org/abs/hep-ph/9508386}{{\tt hep-ph/9508386}}].

\bibitem{Gelis:2001dh}
F.~Gelis and A.~Peshier, {\it {Probing colored glass via q anti-q
  photoproduction. 2. Diffraction}},  {\em Nucl. Phys. A} {\bf 707} (2002)
  175--192, [\href{http://arxiv.org/abs/hep-ph/0111227}{{\tt hep-ph/0111227}}].

\bibitem{Gelis:2001da}
F.~Gelis and A.~Peshier, {\it {Probing colored glass via q anti-q
  photoproduction}},  {\em Nucl. Phys. A} {\bf 697} (2002) 879--901,
  [\href{http://arxiv.org/abs/hep-ph/0107142}{{\tt hep-ph/0107142}}].

\bibitem{Klein:2002wm}
S.~R. Klein, J.~Nystrand, and R.~Vogt, {\it {Heavy quark photoproduction in
  ultraperipheral heavy ion collisions}},  {\em Phys. Rev. C} {\bf 66} (2002)
  044906, [\href{http://arxiv.org/abs/hep-ph/0206220}{{\tt hep-ph/0206220}}].

\bibitem{Armesto:2014sma}
N.~Armesto and A.~H. Rezaeian, {\it {Exclusive vector meson production at high
  energies and gluon saturation}},  {\em Phys. Rev. D} {\bf 90} (2014), no.~5
  054003, [\href{http://arxiv.org/abs/1402.4831}{{\tt arXiv:1402.4831}}].

\bibitem{McLerran:1993ni}
L.~D. McLerran and R.~Venugopalan, {\it {Computing quark and gluon distribution
  functions for very large nuclei}},  {\em Phys. Rev. D} {\bf 49} (1994)
  2233--2241, [\href{http://arxiv.org/abs/hep-ph/9309289}{{\tt
  hep-ph/9309289}}].

\bibitem{McLerran:1993ka}
L.~D. McLerran and R.~Venugopalan, {\it {Gluon distribution functions for very
  large nuclei at small transverse momentum}},  {\em Phys. Rev. D} {\bf 49}
  (1994) 3352--3355, [\href{http://arxiv.org/abs/hep-ph/9311205}{{\tt
  hep-ph/9311205}}].

\bibitem{McLerran:1994vd}
L.~D. McLerran and R.~Venugopalan, {\it {Green's functions in the color field
  of a large nucleus}},  {\em Phys. Rev. D} {\bf 50} (1994) 2225--2233,
  [\href{http://arxiv.org/abs/hep-ph/9402335}{{\tt hep-ph/9402335}}].

\bibitem{Bauer:2000ew}
C.~W. Bauer, S.~Fleming, and M.~E. Luke, {\it {Summing Sudakov logarithms in $B
  \to X_s \gamma $in effective field theory.}},  {\em Phys. Rev. D} {\bf 63}
  (2000) 014006, [\href{http://arxiv.org/abs/hep-ph/0005275}{{\tt
  hep-ph/0005275}}].

\bibitem{Bauer:2000yr}
C.~W. Bauer, S.~Fleming, D.~Pirjol, and I.~W. Stewart, {\it {An Effective field
  theory for collinear and soft gluons: Heavy to light decays}},  {\em Phys.
  Rev. D} {\bf 63} (2001) 114020,
  [\href{http://arxiv.org/abs/hep-ph/0011336}{{\tt hep-ph/0011336}}].

\bibitem{Bauer:2001ct}
C.~W. Bauer and I.~W. Stewart, {\it {Invariant operators in collinear effective
  theory}},  {\em Phys. Lett. B} {\bf 516} (2001) 134--142,
  [\href{http://arxiv.org/abs/hep-ph/0107001}{{\tt hep-ph/0107001}}].

\bibitem{Bauer:2001yt}
C.~W. Bauer, D.~Pirjol, and I.~W. Stewart, {\it {Soft collinear factorization
  in effective field theory}},  {\em Phys. Rev. D} {\bf 65} (2002) 054022,
  [\href{http://arxiv.org/abs/hep-ph/0109045}{{\tt hep-ph/0109045}}].

\bibitem{Bauer:2002nz}
C.~W. Bauer, S.~Fleming, D.~Pirjol, I.~Z. Rothstein, and I.~W. Stewart, {\it
  {Hard scattering factorization from effective field theory}},  {\em Phys.
  Rev. D} {\bf 66} (2002) 014017,
  [\href{http://arxiv.org/abs/hep-ph/0202088}{{\tt hep-ph/0202088}}].

\bibitem{Rothstein:2016bsq}
I.~Z. Rothstein and I.~W. Stewart, {\it {An Effective Field Theory for Forward
  Scattering and Factorization Violation}},  {\em JHEP} {\bf 08} (2016) 025,
  [\href{http://arxiv.org/abs/1601.04695}{{\tt arXiv:1601.04695}}].

\bibitem{Moult:2022lfy}
I.~Moult, S.~Raman, G.~Ridgway, and I.~W. Stewart, {\it {Anomalous dimensions
  from soft Regge constants}},  {\em JHEP} {\bf 05} (2023) 025,
  [\href{http://arxiv.org/abs/2207.02859}{{\tt arXiv:2207.02859}}].

\bibitem{Gao:2024qsg}
A.~Gao, I.~Moult, S.~Raman, G.~Ridgway, and I.~W. Stewart, {\it {A collinear
  perspective on the Regge limit}},  {\em JHEP} {\bf 05} (2024) 328,
  [\href{http://arxiv.org/abs/2401.00931}{{\tt arXiv:2401.00931}}].

\bibitem{Gao:2024fyz}
A.~Gao, I.~Moult, S.~Raman, G.~Ridgway, and I.~W. Stewart, {\it {Reggeization
  in Color}},  \href{http://arxiv.org/abs/2411.09692}{{\tt arXiv:2411.09692}}.

\bibitem{Neill:2023jcd}
D.~Neill, A.~Pathak, and I.~W. Stewart, {\it {Small-x factorization from
  effective field theory}},  {\em JHEP} {\bf 09} (2023) 089,
  [\href{http://arxiv.org/abs/2303.13710}{{\tt arXiv:2303.13710}}].

\bibitem{Ee:2026cvm}
J.-H. Ee, C.~Lee, and S.~T. Schindler, {\it {Minding the gap between hard and
  diffractive structure functions at the EIC}},
  \href{http://arxiv.org/abs/2607.12006}{{\tt arXiv:2607.12006}}.

\bibitem{vonWeizsacker:1934nji}
C.~F. von Weizsacker, {\it {Radiation emitted in collisions of very fast
  electrons}},  {\em Z. Phys.} {\bf 88} (1934) 612--625.

\bibitem{Williams:1934ad}
E.~J. Williams, {\it {Nature of the high-energy particles of penetrating
  radiation and status of ionization and radiation formulae}},  {\em Phys.
  Rev.} {\bf 45} (1934) 729--730.

\bibitem{Frixione:1993yw}
S.~Frixione, M.~L. Mangano, P.~Nason, and G.~Ridolfi, {\it {Improving the
  Weizsacker-Williams approximation in electron - proton collisions}},  {\em
  Phys. Lett. B} {\bf 319} (1993) 339--345,
  [\href{http://arxiv.org/abs/hep-ph/9310350}{{\tt hep-ph/9310350}}].

\bibitem{Muller:2012zq}
B.~Muller, J.~Schukraft, and B.~Wyslouch, {\it {First Results from Pb+Pb
  collisions at the LHC}},  {\em Ann. Rev. Nucl. Part. Sci.} {\bf 62} (2012)
  361--386, [\href{http://arxiv.org/abs/1202.3233}{{\tt arXiv:1202.3233}}].

\bibitem{STAR:2015mki}
{\bf STAR} Collaboration, L.~Adamczyk et~al., {\it {Azimuthal anisotropy in
  U$+$U and Au$+$Au collisions at RHIC}},  {\em Phys. Rev. Lett.} {\bf 115}
  (2015), no.~22 222301, [\href{http://arxiv.org/abs/1505.07812}{{\tt
  arXiv:1505.07812}}].

\bibitem{Klein:2016yzr}
S.~R. Klein, J.~Nystrand, J.~Seger, Y.~Gorbunov, and J.~Butterworth, {\it
  {STARlight: A Monte Carlo simulation program for ultra-peripheral collisions
  of relativistic ions}},  {\em Comput. Phys. Commun.} {\bf 212} (2017)
  258--268, [\href{http://arxiv.org/abs/1607.03838}{{\tt arXiv:1607.03838}}].

\bibitem{ALICE:2014eof}
{\bf ALICE} Collaboration, B.~B. Abelev et~al., {\it {Exclusive
  $\mathrm{J/}\psi$ photoproduction off protons in ultra-peripheral p-Pb
  collisions at $\sqrt{s_{\rm NN}}=5.02$ TeV}},  {\em Phys. Rev. Lett.} {\bf
  113} (2014), no.~23 232504, [\href{http://arxiv.org/abs/1406.7819}{{\tt
  arXiv:1406.7819}}].

\bibitem{ALICE:2015mzu}
{\bf ALICE} Collaboration, J.~Adam et~al., {\it {Measurement of an excess in
  the yield of $J/\psi$ at very low $p_{\rm T}$ in Pb-Pb collisions at
  $\sqrt{s_{\rm NN}}$ = 2.76 TeV}},  {\em Phys. Rev. Lett.} {\bf 116} (2016),
  no.~22 222301, [\href{http://arxiv.org/abs/1509.08802}{{\tt
  arXiv:1509.08802}}].

\bibitem{STAR:2023nos}
{\bf STAR} Collaboration, M.~I. Abdulhamid et~al., {\it {Observation of Strong
  Nuclear Suppression in Exclusive J/{\ensuremath{\psi}} Photoproduction in
  Au+Au Ultraperipheral Collisions at RHIC}},  {\em Phys. Rev. Lett.} {\bf 133}
  (2024), no.~5 052301, [\href{http://arxiv.org/abs/2311.13637}{{\tt
  arXiv:2311.13637}}].

\bibitem{STAR:2025wpi}
{\bf STAR} Collaboration, B.~E. Aboona et~al., {\it {Evidence of
  Spin-Interference Effects in Exclusive
  J/{\ensuremath{\psi}}{\textrightarrow}e+e- Photoproduction in Ultraperipheral
  Heavy-Ion Collisions}},  {\em Phys. Rev. Lett.} {\bf 136} (2026), no.~24
  242302, [\href{http://arxiv.org/abs/2512.02865}{{\tt arXiv:2512.02865}}].

\bibitem{CMS:2025lsm}
{\bf CMS} Collaboration, V.~Chekhovsky et~al., {\it {Observation of Coherent
  {\ensuremath{\phi}}(1020) Meson Photoproduction in Ultraperipheral PbPb
  Collisions at sNN=5.36{\,}{\,}TeV}},  {\em Phys. Rev. Lett.} {\bf 135}
  (2025), no.~26 262301, [\href{http://arxiv.org/abs/2504.05193}{{\tt
  arXiv:2504.05193}}].

\bibitem{CMS:2025oxg}
{\bf CMS} Collaboration, V.~Chekhovsky et~al., {\it {Probing Gluon Fluctuations
  in Nuclei with the First Energy-Dependent Measurement of Incoherent
  J/{\ensuremath{\psi}} Photoproduction in Ultraperipheral PbPb Collisions}},
  {\em Phys. Rev. Lett.} {\bf 135} (2025), no.~11 112301,
  [\href{http://arxiv.org/abs/2503.08903}{{\tt arXiv:2503.08903}}].

\bibitem{CMS:2026soy}
{\bf CMS} Collaboration, A.~Belyaev et~al., {\it {Observation of Nuclear
  Suppression in Coherent {\ensuremath{\Upsilon}}(1S) Photoproduction off Heavy
  Nuclei at the LHC}},  {\em Phys. Rev. Lett.} {\bf 137} (2026), no.~14 142301,
  [\href{http://arxiv.org/abs/2604.05814}{{\tt arXiv:2604.05814}}].

\bibitem{ATLAS:2026cxr}
{\bf ATLAS} Collaboration, G.~Aad et~al., {\it {Measurement of jet
  photoproduction in ultra-peripheral Pb+Pb collisions without nuclear breakup
  at $\sqrt{s_\mathrm{NN}} = 5.02$ TeV with the ATLAS detector}},
  \href{http://arxiv.org/abs/2604.24435}{{\tt arXiv:2604.24435}}.

\bibitem{ATLAS:2026sct}
{\bf ATLAS} Collaboration, G.~Aad et~al., {\it {Observation of impact parameter
  dependent modifications of nuclear parton distributions in photonuclear Pb+Pb
  collisions at $\sqrt{s_\mathrm{NN}} = 5.02$ TeV with the ATLAS detector}},
  \href{http://arxiv.org/abs/2604.20559}{{\tt arXiv:2604.20559}}.

\bibitem{Miller:2007ri}
M.~L. Miller, K.~Reygers, S.~J. Sanders, and P.~Steinberg, {\it {Glauber
  modeling in high energy nuclear collisions}},  {\em Ann. Rev. Nucl. Part.
  Sci.} {\bf 57} (2007) 205--243,
  [\href{http://arxiv.org/abs/nucl-ex/0701025}{{\tt nucl-ex/0701025}}].

\bibitem{Baltz:2002pp}
A.~J. Baltz, S.~R. Klein, and J.~Nystrand, {\it {Coherent vector meson
  photoproduction with nuclear breakup in relativistic heavy ion collisions}},
  {\em Phys. Rev. Lett.} {\bf 89} (2002) 012301,
  [\href{http://arxiv.org/abs/nucl-th/0205031}{{\tt nucl-th/0205031}}].

\bibitem{Cacciari:2025tgr}
M.~Cacciari, G.~M. Innocenti, and A.~M. Sta{\'s}to, {\it {Inclusive open charm
  photoproduction in ultraperipheral collisions at the LHC with in the
  generalized photon-nucleus fixed-order next-to-leading logarithm framework}},
   {\em Phys. Rev. D} {\bf 112} (2025), no.~9 094029,
  [\href{http://arxiv.org/abs/2506.09893}{{\tt arXiv:2506.09893}}].

\bibitem{Klasen:2002xb}
M.~Klasen, {\it {Theory of hard photoproduction}},  {\em Rev. Mod. Phys.} {\bf
  74} (2002) 1221--1282, [\href{http://arxiv.org/abs/hep-ph/0206169}{{\tt
  hep-ph/0206169}}].

\bibitem{Klasen:2004qr}
M.~Klasen and G.~Kramer, {\it {Factorization breaking in diffractive dijet
  photoproduction}},  {\em Eur. Phys. J. C} {\bf 38} (2004) 93--104,
  [\href{http://arxiv.org/abs/hep-ph/0408203}{{\tt hep-ph/0408203}}].

\bibitem{Brodsky:1971ud}
S.~J. Brodsky, T.~Kinoshita, and H.~Terazawa, {\it {Two Photon Mechanism of
  Particle Production by High-Energy Colliding Beams}},  {\em Phys. Rev. D}
  {\bf 4} (1971) 1532--1557.

\bibitem{Budnev:1975poe}
V.~M. Budnev, I.~F. Ginzburg, G.~V. Meledin, and V.~G. Serbo, {\it {The Two
  photon particle production mechanism. Physical problems. Applications.
  Equivalent photon approximation}},  {\em Phys. Rept.} {\bf 15} (1975)
  181--281.

\bibitem{ATLAS:2017fur}
{\bf ATLAS} Collaboration, M.~Aaboud et~al., {\it {Evidence for light-by-light
  scattering in heavy-ion collisions with the ATLAS detector at the LHC}},
  {\em Nature Phys.} {\bf 13} (2017), no.~9 852--858,
  [\href{http://arxiv.org/abs/1702.01625}{{\tt arXiv:1702.01625}}].

\bibitem{CMS:2018erd}
{\bf CMS} Collaboration, A.~M. Sirunyan et~al., {\it {Evidence for
  light-by-light scattering and searches for axion-like particles in
  ultraperipheral PbPb collisions at $\sqrt{s_\mathrm{NN}} =$ 5.02 TeV}},  {\em
  Phys. Lett. B} {\bf 797} (2019) 134826,
  [\href{http://arxiv.org/abs/1810.04602}{{\tt arXiv:1810.04602}}].

\bibitem{Cavasinni:1985bi}
V.~Cavasinni, T.~Del~Prete, M.~Morganti, F.~Schiavo, D.~Lloyd~Owen,
  G.~Paternoster, S.~Patricelli, and G.~Anzivino, {\it {Observation of Double
  Pomeron Exchange in $\alpha^- \alpha$ Collisions at the {CERN} Intersecting
  Storage Rings}},  {\em Z. Phys. C} {\bf 28} (1985) 487.

\bibitem{Roldao:2000ze}
C.~G. Roldao and A.~A. Natale, {\it {Photon-photon and pomeron-pomeron
  processes in peripheral heavy ion collisions}},  {\em Phys. Rev. C} {\bf 61}
  (2000) 064907, [\href{http://arxiv.org/abs/nucl-th/0003038}{{\tt
  nucl-th/0003038}}].

\bibitem{Albrow:2017gle}
M.~Albrow, {\it {Hadron Spectroscopy in Double Pomeron Exchange Experiments}},
  {\em AIP Conf. Proc.} {\bf 1819} (2017), no.~1 040008,
  [\href{http://arxiv.org/abs/1701.09092}{{\tt arXiv:1701.09092}}].

\bibitem{CMS:2023lfr}
{\bf CMS} Collaboration, A.~Tumasyan et~al., {\it {First Measurement of the
  Forward Rapidity Gap Distribution in pPb Collisions at sNN=8.16{\,}{\,}TeV}},
   {\em Phys. Rev. D} {\bf 108} (2023), no.~9 092004,
  [\href{http://arxiv.org/abs/2301.07630}{{\tt arXiv:2301.07630}}].

\end{thebibliography}\endgroup

\end{document}